\documentclass[final,3p,times]{elsarticle}
\usepackage{graphicx}
\usepackage[utf8]{inputenc}
\usepackage[T1]{fontenc}
\usepackage{amsmath}
\usepackage[varg]{newtxmath} 
\usepackage[dvipsnames]{xcolor}
\usepackage{pgfplots, tikz}
\usepackage{subfigure, dcolumn, bm, color, accents, hyperref, tikz ,booktabs}
\hypersetup{
    colorlinks   = true,
    citecolor    = blue,
    linkcolor    = blue}

\usepackage{doi}

\title{An explicit multiscale pseudo orbit-averaging time integration algorithm}

\author[1,2]{M. H. Rosen\corref{cor1}}
\ead{mhrosen@pppl.gov}

\author[2]{M. Francisquez}
\author[1,2]{G. W. Hammett}

\cortext[cor1]{Corresponding author: email address mhrosen@pppl.gov}

\affiliation[1]{organization={Department of Astrophysical Sciences, Princeton University},
city={Princeton},
postcode={08540},
state={NJ},
country={USA}}

\affiliation[2]{organization={Princeton Plasma Physics Laboratory},
city={Princeton},
postcode={08540},
state={NJ},
country={USA}}

\begin{document}
\begin{abstract}

We present an explicit multiscale algorithm for solving differential equations for problems with high-frequency modes that can be averaged over by separating and scaling the fast and slow dynamics within a single equation.
We introduce a phased time integrator for cases where the boundaries of dynamical scales are known: one phase solves the unmodified equation, while the other freezes part of phase-space and slows down the evolution of the fast dynamics. This algorithm is applied to reduced kinetic models of plasmas in magnetic mirrors, which feature a distinct boundary between a region dominated by rapid particle transit and a region characterized by slow collisions. Two representative model problems are presented that decompose the dynamics of the magnetic mirror into a simpler, computationally inexpensive form. 
The model problems demonstrate a speedup by a factor of order $\omega / \nu_c$, where $\omega$ is the fast oscillation frequency and $\nu_c$ is the slow damping rate. This is a 30,000$\times$ speedup for a case of practical interest.
\end{abstract}

\maketitle

\tableofcontents

\section{Introduction}
\label{sec: introduction}

Numerous systems in physics, engineering and applied mathematics involve the coexistence of fast, often periodic, motion and much slower processes that determine the steady-state equilibrium. For instance, a rapidly oscillating damped pendulum with a slow evolving envelope, orbital galactic and stellar dynamics, slowly reacting but fast-moving molecular dynamics, or disentangling the fast and slow evolution of loss functions in optimizing neural networks \cite{liu2025neural, schlick2010molecular, binney2011galactic, abreu2021multiple}. 

Explicit integration techniques impose a Courant condition, or time-step stability limit, determined by the fastest dynamics for the entire simulation, so evolving the equations on the long time scale of the slow dynamics can be prohibitively expensive. Implicit integration techniques bypass the Courant condition; however, they can be challenging to implement and expensive due to the complexity and computational cost of the required matrix inversions. Iterative methods for solving the resulting large, sparse matrix equations can sometimes require many iterations. Iterative solvers that work well for elliptic or parabolic (diffusion) operators (which lead to symmetric positive definite matrices) are often not very effective for advection-dominated problems (hyperbolic-type problems leading to non-symmetric matrices with large eigenvalues near the imaginary axis). Although iterative methods 
have been developed that are relatively efficient for some advection-dominated problems, even matrix-free iterative methods still require a large number of global reduction operations per time step, unlike explicit methods \citep{Chacon:2003, Knoll:2004, Knoll:2005, Loureiro-Hammett:2008}. This difference can significantly hinder the performance of implicit methods on GPU-based or very large-scale distributed-memory systems.
There have been interesting recent advances in multigrid methods for advection-dominated problems using variations of Approximate Ideal Restriction (AIR) \citep{Manteuffel:2018, Manteuffel:2019} combined with certain preconditioners that have been effective for gyrokinetic problems \citep{Dorf:2025}.
Yet, there is continued motivation to explore explicit integrators that are potentially faster or simpler, especially for different types of problems or numerical methods. Here, we explore one such algorithm exploiting the scale separation in the system to take larger time steps while retaining accuracy.

This work pertains to differential equations which can be split into fast and slow operators
\begin{align}
\frac{\partial f}{\partial t} 
 = \mathcal{L}(f) = \mathcal{L}_{\mathrm{fast}}(f) + \mathcal{L}_{\mathrm{slow}}(f). \label{eq: Fast-slow split example introduction}
\end{align}
We write $f$ generically: it may be a field $f(\vec x,t)$ or a
vector $f_i(t)$.
The pseudo orbit-averaging (POA) algorithm we describe here recognizes two distinct splittings. One split is between the fast and slow operators,
and another is between different regions in $\vec x$ space where the fast operator has different behaviors, leading to fast nearly-periodic orbits that can be averaged over in one region, and fast transit and decay in the other.

\begin{figure}
    \centering\includegraphics[width=0.9\linewidth]{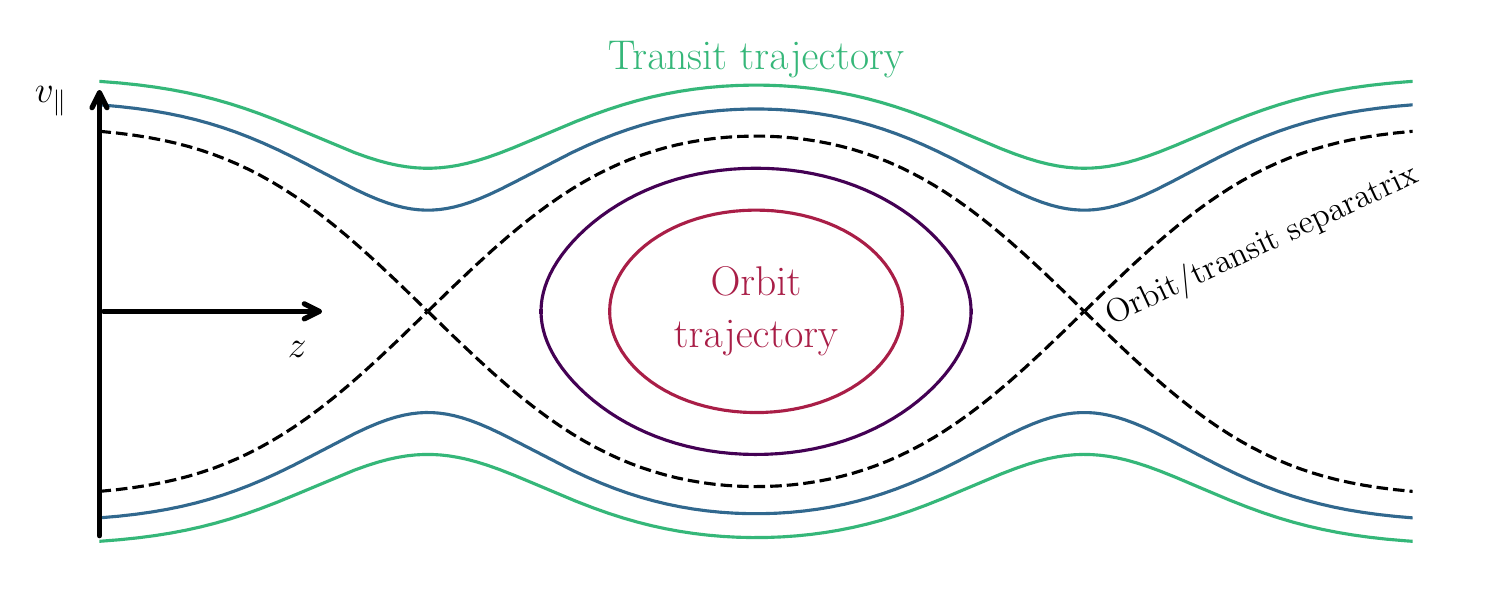}
    \caption{An illustration of the differences between trajectories in phase space for particles on transit trajectories versus orbit trajectories. Orbiting trajectories feature a periodic motion, while transiting trajectories do not and will be lost at a boundary in $z$. Here, $z$ is a spatial coordinate and $v_\parallel$ is a 1D velocity, $\dot z$.}
    \label{fig: orbits versus transit trajectory}
\end{figure}

For instance, consider a kinetic problem where collisions are rare, some particles orbit periodically, and others transit freely and are quickly lost. These two regions of phase space, illustrated in figure \ref{fig: orbits versus transit trajectory}, are sometimes called the trapped and passing regions, or here, the orbit and transit regions. The average density in the orbit region evolves only slowly due to weak collisions that scatter particles from periodic orbits onto trajectories that are quickly lost. We would like to advance that averaged component in time with a large time step. Particles in the transit region are quickly lost as they flow out of the system, so this region must be advanced with a small time step. 
In effect, the POA algorithm we present here evolves different components of $f$ at different rates in different stages of the algorithm.

In plasma physics, the idea of averaging over fast, periodic, oscillatory motion is pervasive, beginning with the early work by Alfv\'en, Northrop, and others on the slow drifts of the guiding centers of particles after averaging over their fast cyclotron motion and later extended to gyrokinetic equations. In the context of magnetic mirrors (and neoclassical theory and other applications in tokamaks and stellarators), analytical studies have long used the technique of the bounce-average to step over the fast advective orbit motion in these problems \citep{bendaniel1962scattering, marx1970effects, Pastukhov_1974, Mauel:1984, Hammett:1986, matsuda1986relativistic, harvey1992cql3d, li2025approximating, Rosen_Sengupta_Ochs_Parra_Hammett_2025}, and a related numerical ``goosing'' technique has been used to accelerate Monte Carlo simulations of beam injection in plasmas\cite{Goldston:1981:JCP}. A comprehensive overview of the technique is presented in \citet{Ochs_2025}, where the authors discuss the application of bounce-average techniques to multi-well plasmas. Inspired by the bounce-average, this article introduces the POA algorithm, a time integration technique that modifies the original equation. To find the steady-state equilibrium solution to equation \eqref{eq: Fast-slow split example introduction}, this article proposes a two-phase integrator consisting of the full-dynamics phase (FDP) and the (pseudo) orbit-averaged phase (OAP). 
During the FDP, equation \eqref{eq: Fast-slow split example introduction} is solved directly. During an OAP, $\mathcal{L}_{\mathrm{fast}}$ is replaced by $\alpha\,\mathcal{L}_{\mathrm{fast}}$ (where $\alpha \ll 1$ when the orbit frequency is very fast compared to the slow dynamics), and we freeze the transiting region by applying a mask $H_{\rm orbit}$, so that only orbiting particles are evolved:
\begin{equation} 
\frac{\partial f}{\partial t} = H_{\rm orbit} \left( \alpha\mathcal{L}_{\mathrm{fast}}(f) + {\mathcal{L}}_{\mathrm{slow}}(f)\right). \label{eq: orbit average split example introduction}
\end{equation}
where $H_{\rm orbit} =1$ in the orbiting part of phase space, while $H_{\rm orbit} = 0$ in the transiting region.
Because the fast term is slowed down in the OAP by the factor $\alpha$, we can take larger time steps to reach equilibrium more quickly in the orbiting region. The OAP is alternated with FDP to reach equilibrium.
A practical detail in analytical orbit-averaging is the calculation of the bounce-averaged collision operator and transformation to coordinates that are (nearly) constants of the motion.
The POA algorithm circumvents this, making it a simpler technique that existing explicit solvers can readily employ.

The POA algorithm is inspired by and borrows ideas from several established multiscale time-integration frameworks, such as equation-free modeling and projective integration, Heterogeneous Multiscale Methods (HMMs), and multirate methods.
To the best of our knowledge, it has unique features useful for the class of problems we consider and is a one-of-a-kind algorithm for accelerating the calculation of steady-state equilibria.

In particular, we target problems in which a single kinetic equation contains phase-space regions dominated by dynamics with different properties (one region with fast oscillatory dynamics with very little damping, and another region with fast dynamics that are strongly damped), and in which it is comparatively straightforward to modify selected terms in the governing operator. This leads to an explicit multiscale integrator that can be implemented with minimal changes to existing solvers, while still retaining the ability to recover the full fast dynamics during corrective phases. 

Equation-free modeling with projective-integration provides a paradigm for accelerating simulations when one expects that suitably chosen macroscopic variables could be described by macroscopic equations with slow dynamics, but the actual form of those equations is not necessarily known (see \cite{Theodoropoulos2000,Gear2002CoarseIntegration, Gear_Kevrekidis_2003a} and the review in \cite{Kevrekidis2009Review, Kevrekidis2010Scholarpedia}).
Using projective integration, one typically takes a short burst of fine-scale (inner) steps with the full microscopic equations to damp fast transients and estimate the time derivative of the slow evolution, followed by a larger projective (outer) step that advances the coarse macroscopic variables.
This approach has been successfully applied in a variety of contexts, including kinetic problems \citep{Bailo_Rey_2022}. 

Similarly, HMMs are a related general framework for multiscale computation in which a macroscale solver is coupled to localized microscale simulations that provide missing constitutive information (e.g., fluxes, effective parameters, or closures) needed to advance the macroscopic model \citep{cms/1118150402}. An extensive review is given by \citet{Heterogenous_multiscale_methods_a_review_2007}, and the stiff-ODE setting has been analyzed in detail by \citet{Engquist_Tsai_2005}. HMM is particularly attractive when one can clearly identify appropriate coarse variables and a macroscale evolution law (possibly up to missing closure terms).
Furthermore, the POA scheme also shares motivation with inter/extrapolation-based multirate methods, by advancing slow collisional dynamics without being restricted by rapid transit times, but the mechanism differs. In multirate methods, the system is partitioned into slow and fast subsystems (or additive components), each integrated on its own time grid, with the missing variables reconstructed via interpolation or extrapolation (for example, spline-based coupling) \citep{gear1984multirate, constantinescu2007multirate, bartel2020inter, schafers2023spline, gunther2025multirate}.
Instead of isolating macroscopic variables with slow dynamics, the POA modifies the governing equation by splitting phase space into two regions (orbiting and transit), and suppressing the fast dynamics that are still present in the orbiting region.

The POA algorithm bears similarities to recent work on time-dilation, multizone, and zoom-in types of multiscale methods originally developed for GRMHD problems in astrophysics \cite{Rosen2025aps-poster, hopkins2025time, Cho2024zone, Guo2025zoom}, but that can have broader applicability.
There are also some similarities with an acceleration method for PIC \cite{werner2018speeding} codes.
In a sense, POA can be viewed as a phase-space-selective and term-selective extension of time-dilation and related ideas, with a natural corrective mechanism through alternation of modified and full-dynamics phases, particularly useful for problems with identifiable orbiting and transiting regions.
The time-dilation introduced by the POA algorithm affects the temporal accuracy of the solution; however, in equilibrium, temporal accuracy is irrelevant.
Assessing applications that require accurate temporal tracking would benefit from additional analysis of how the operator splitting and equation modification approach affect overall temporal accuracy; this is left for future work.

We demonstrate the POA integrator using reduced models inspired by kinetic ones for plasmas in magnetic mirrors
\citep{Francisquez_Rosen_Mandell_Hakim_Forest_Hammett_2023, Tran_Frank_Le_Stanier_Wetherton_Egedal_Endrizzi_Harvey_Petrov_Qian_et, Frank_Viola_Petrov_Anderson_Bindl_Biswas_Caneses_Endrizzi_Furlong_Harvey_et}. In this problem, the distribution function evolves due to fast advection, weak collisions, and a source. Phase space is partitioned into regions of collision-dominated orbiting particles and advection-dominated passing particles. The POA algorithm exploits orbit-average symmetry, with the OAP resembling the orbit-averaged equation.

This article describes the POA algorithm and its use in reduced models in the following order. Section \ref{sec: algorithm} describes the algorithm. Section \ref{sec: 1D model example} explores a model that describes the velocity diffusion of electrostatically confined particles into the transit region. Section \ref{sec: five equation toy model} explores an alternative approach to capturing the orbiting advection of particles, including the treatment of exhaust in the transit region. This section contains relevant variations of the second model, moving around the source and orbit/transit coupling.
Section \ref{sec: conclusion} offers concluding remarks.
\section{The Pseudo Orbit-Averaging (POA) Algorithm}
\label{sec: algorithm}

\subsection{Toy Problem Motivation: Orbit Averaging}
\label{subsec:toy-OA}

To motivate the pseudo orbit-averaging algorithm, we first review standard orbit averaging (in some fields, this technique is called bounce-averaging) in a simplified problem.
We will consider the following PDE as a simplified model of the phase space dynamics in figure \ref{fig: orbits versus transit trajectory},
\begin{align}
\frac{\partial f(\theta,v,t)}{\partial t} = - \Omega \frac{\partial f}{\partial \theta} + \frac{\partial}{\partial v} \left(  D(\theta,v) \frac{\partial f}{\partial v} \right)  + S(\theta,v)
\label{eq:toy-OA}
\end{align}
Here, $f$ is the density of particles in phase space, $\theta$ is an angle-like coordinate that measures where along a trajectory (a surface of constant energy) a particle is, and $v$ labels which trajectory a particle is on. (The coordinate $v$ here is a constant of the collisionless motion and could be the velocity on a trajectory surface. 
One could develop a more precise mapping from $(z,v_\parallel)$ phase-space coordinates to the $(\theta,v)$ coordinates used in this PDE, but this qualitative relationship is sufficient as a simplified model.)

In this PDE, the first term on the RHS represents advection around a trajectory at frequency $\Omega > 0$, the second term represents diffusion of particles in the $v$ coordinates, and $S$ is a source term.
Particles with $0<v<a$ are trapped and are on trajectories that execute closed orbits. Particles with $a<v<1$ are in the unconfined region and are on open trajectories that will carry them out of the system to be lost at boundaries in $\theta$.
We will model this with periodic boundary conditions $f(-\pi,v,t) = f(\pi,v,t)$ for the orbit region $0<v<a<1$, but there are no incoming particles in the transit region $a<v<1$ where the boundary condition is $f(-\pi,v,t) = 0$.
For the diffusion term, the boundary conditions in $v$ will be $\partial f/\partial v = 0$ at $v=0$ (i.e., conservative) and $f = 0 $ at $v = 1$.

Equation \eqref{eq:toy-OA} can be a model for many types of physical systems. It could be for electrons trapped in an electrostatic well (which might be produced in a magnetic mirror), injected at low energy and bouncing back and forth until they diffuse via collisions to higher energies, where they are lost.
It could be for ions in a magnetic mirror, where $v$ is related to the pitch angle.
Alternatively, we could change $v$ to $r$ and think of this as a model of particle transport in a tokamak, where particles circulate in poloidal angle $\theta$ many times before they diffuse radially to the plasma edge to be lost.

We are considering the case where the orbit frequency $\Omega$ is large compared to all of the other terms, which scale with a typical diffusion rate $\nu \sim D/a^2 $, 
so particles are going around closed orbit trajectories many times before they slowly diffuse out to a transit trajectory, where they are quickly lost.
Defining an expansion parameter $\epsilon = \nu/\Omega \rightarrow 0$, we will do an asymptotic expansion treating the $\Omega$ term in \eqref{eq:toy-OA} as $1/\epsilon$ larger than all of the other terms. Expanding $f = f_0 + f_1 + \ldots$ (where $f_1 \sim \epsilon f_0$), to lowest order we have $0 = df_0 /d\theta$
so $f_0$ must be constant along $\theta$. To next order, we have
\begin{align}
\frac{\partial f_0}{\partial t} = - \Omega \frac{\partial f_1}{\partial \theta} + 
\frac{\partial}{\partial v} \left( D \frac{\partial f_0}{\partial v} \right)
 + S    
\end{align}
The solution for $f_0$ seems to require knowing $f_1$, but this term can be annihilated by applying an orbit average operator that averages over a trajectory:
\begin{align}
   \langle \ldots \rangle = \frac{1}{T} \int d\tau \ldots
         =  \frac{1}{2 \pi} \int_{-\pi}^\pi d\theta  \ldots   
\end{align}
In the orbit region we can use $\int d \theta \, \partial f_1 / \partial \theta = 0$ from periodicity so that this leads to:
\begin{align}
   \frac{\partial f_0}{\partial t} =
\frac{\partial}{\partial v} \left( 
\langle D \rangle \frac{\partial f_0}{\partial v} \right)
 + \langle S \rangle
\end{align}
Notice that the orbit-averaged equation for $f_0$ looks almost identical to the original equation \ref{eq:toy-OA}, but just drops the fast operator (and replaces some of the coefficients and source term with their orbit averages).
The resulting $f_0$ is independent of $\Omega$ in this limit, so the exact value of $\Omega$ does not matter as long as it is fast compared to the characteristic damping rate of the diffusion term $\nu\sim D/a^2$.
Because of this, we can multiply the $\Omega$ term by some small coefficient $\alpha$ to be computationally much quicker while still getting approximately the same answer for $f$ as long as $\alpha \Omega$ is still large compared to $\nu$.
This is the main inspiration for the POA algorithm.

Many standard bounce-averaged theories and codes stop at this order. The boundary condition used is then that $f_0(v,t) = 0$ at $v=a$, since $f_0$ is independent of $\theta$ and the boundary condition is $f(\theta=-\pi,v,t) = 0$ for $v\ge a$.
Even though $f_0$ = 0 in the transit region in this approximation, one can still calculate the net flux of particles $\Gamma$ into the transit region that will be lost, using $\Gamma = - \langle D \rangle \partial f_0/\partial v|_{v=a}$.

However, in some problems, one seeks to determine the detailed structure of the solution in the transit region near $v = a$ with more accuracy.
For example, it is important to determine the scrape-off-layer heat-flux width to the divertor plates in tokamaks and mirrors.
This requires a modification to the above asymptotic expansion, which breaks down near $v = a$ where there is a narrow boundary layer of width $\Delta$ so that the diffusion term $\sim D \partial^2 f/\partial v^2 \sim D f / \Delta^2$ becomes comparable to the fast transit term $\alpha \Omega f$. This leads to a boundary layer width $\Delta \sim \sqrt{D/ (\alpha \Omega)} \sim a \sqrt{\epsilon/\alpha}$.
Thus, the actual ratio of the fast to slow terms matters here.
For this reason, the POA algorithm will alternate between two stages, one with $\alpha \sim \nu/\Omega \ll 1 $ for faster computation of the solution in the orbiting region $v<a$, and one with the physical value of $\alpha=1$ for the full domain including the transit region.
Because the ratio of the fast terms to other terms, $\alpha \Omega / \nu$, is artificially reduced in the OAP, oscillations from $f_1$ can be enhanced in some cases, depending on how strong the $\theta$ variation of $S$ or $D$ is (or their equivalent in more complicated problems).

\subsection{The General POA Algorithm}
\label{sec:POA-algorithm}

Consider a kinetic equation of the form:
\begin{equation}
\frac{\partial f}{\partial t} = \{\mathcal{H},f\} + C(f) + S, \label{eq: Full vlasov equation}
\end{equation}
where $\{\mathcal{H},f\}$ represents the Poisson bracket of the Hamiltonian $\mathcal{H}$ with distribution function $f$, $C(f)$ is a collision operator, and $S$ denotes a source term.
The Poisson bracket $\{\mathcal{H},f\}$ is an advective term in phase space that contains the evolution of both the orbiting and transiting particles. 
The Vlasov-Boltzmann kinetic equation used for gases and plasmas can be written in the form of \eqref{eq: Full vlasov equation}.
In a low collision frequency regime, the Courant limit is set by the fast advection of $\{\mathcal{H},f\}$ and collisions $C(f)$ are slow in comparison.
The algorithm iteratively advances the solution forward in time in two phases:

\begin{enumerate}
\item \textbf{Full dynamics phase (FDP):} \\
Run equation \eqref{eq: Full vlasov equation} with small timesteps $\Delta t_{\rm FDP}$ (as required to satisfy the Courant limit), for a few transit times for a total time $\tau_{\rm FDP}$. This allows the parts of the distribution function on transiting trajectories to evolve towards equilibrium, and also smooths out the solution along advective characteristics in the orbiting region. This phase allows the transit region to relax to a slowly changing state, at which the next phase is executed.

\item \textbf{Orbit-averaged phase (OAP):} \\
$\partial _t f$ is separated into regions of orbits and transits using a step function.
\begin{equation}
\frac{\partial f}{\partial t} = H_{\text{orbit}}\left(\alpha\{\mathcal{H},f\} + C(f) + S\right),
\label{eq: dfdt reduced dynamics frozen fast phase}
\end{equation}
where
\begin{equation}
H_{\text{orbit}} = 
\begin{cases}
1, & \text{in the orbit region},\\[5pt]
0, & \text{in the transit region}.
\end{cases} \label{eq: trapped vs transit step function}
\end{equation}
Here, $\alpha$ is the amount that distribution function evolution is slowed down. 
Orbiting regions of phase space are defined as trajectories that close on themselves, forming a periodic motion, while transit regions are not periodic. Figure \ref{fig: orbits versus transit trajectory} illustrates how orbit regions differ from transit trajectories to show how these regions of phase space have different dynamics.
Equation \eqref{eq: dfdt reduced dynamics frozen fast phase} is advanced with large timesteps $\Delta t_{\rm OAP}$ and $\alpha < 1$ to arrive at collision scale times for a total time $\tau_{\rm OAP}$. By freezing the dynamics of the transiting region of the distribution function, this algorithm allows the orbiting particles to evolve on the slower time scale of collisions while approximating the dynamics in the transiting region.
\end{enumerate}

The theoretical speed-up from this algorithm is a ratio of time saved in each phase, which is $(\tau_{ \rm FDP} + \tau_{\rm OAP} )/ (\tau_{\rm FDP} + \alpha\tau_{\rm OAP})$. In the case where $\tau_{\rm FDP} \ll \tau_{\rm OAP}$, then the speedup is approximately  $1/\alpha$.

\section{One-Dimensional PDE Model}
\label{sec: 1D model example}
This section illustrates the use of the POA algorithm and examines boundary layer effects using a 1D PDE model with slow diffusion in one region and rapid loss in another.

\subsection{Problem Setup} \label{sec: 1D model problem setup and motivation}
Inspired by problems concerning electrostatically confined particles in an adiabatic trap, an ad hoc model kinetic equation is constructed with a low-energy source and a high-energy sink, similar to \citet{Pastukhov_1974} and \citet{Rosen_Sengupta_Ochs_Parra_Hammett_2025}. Collisions are represented by a simple diffusion term, further explained by the Wiener-Hopf problem constructed for mirror machines in \citet{Baldwin_Cordey_Watson_1972}:
\begin{equation}
\frac{\partial f(v,t)}{\partial t} + \nu_{\parallel} H(v-a)f(v,t) = D\frac{\partial^2 f(v,t)}{\partial v^2} + S \,H(a-v) , \label{eq: 1D toy model equation dfdt}
\end{equation}
where $D$ is a diffusion coefficient, $\nu_{\parallel}$ is a parallel streaming rate, and $H(v-a)$ is the Heaviside step function centered at $v=a$, which is the orbit/transit separatrix. 
This equation is a 1D simplification of the 2D model in equation \ref{eq:toy-OA}, dropping the $\Omega \, \partial / \partial \theta$ term in the orbit region ($v<a$) and replacing it by a fast loss at the rate $\nu_\parallel$ in the transit region $v>a$.
This equation models the balance between three processes: a uniform source term $S \, H(a-v)$ that fuels particles only within the orbiting region ($v < a$), diffusive collisions represented by $D \partial^2 f/\partial v^2$, and the advective loss of transiting particles through the sink term $-\nu_{\parallel}H(v-a)f$.

Equation \eqref{eq: 1D toy model equation dfdt} accepts a straightforward solution in equilibrium ($\partial f/\partial t = 0$). For particles in the orbiting region ($v < a$), the balance between diffusion and the uniform source yields a parabolic profile:
\begin{align}
D\frac{\partial f}{\partial v} &= -\int_0^v S\,dv = -vS,\\[5pt]
f(v) &= f_0 - \frac{v^2 S}{2D}.\label{eq: 1D toy solution v<a}
\end{align}
In the transit region ($v > a$), particles experience exponential decay due to the sink:
\begin{align}
D\frac{\partial^2 f}{\partial v^2} &= \nu_{\parallel}f,\\[5pt]
f(v) &= \frac{aS}{\sqrt{D \nu_\parallel}}\,e^{-\sqrt{\frac{\nu_{\parallel}}{D}}(v-a)}. \label{eq: 1D toy solution v>a}
\end{align}
A matching condition at $v=a$ ensures continuity of the solution across the orbit/transit separatrix and defines the coefficient $f_0$.

\begin{figure}[t]
    \centering
    \includegraphics[width=0.6\linewidth]{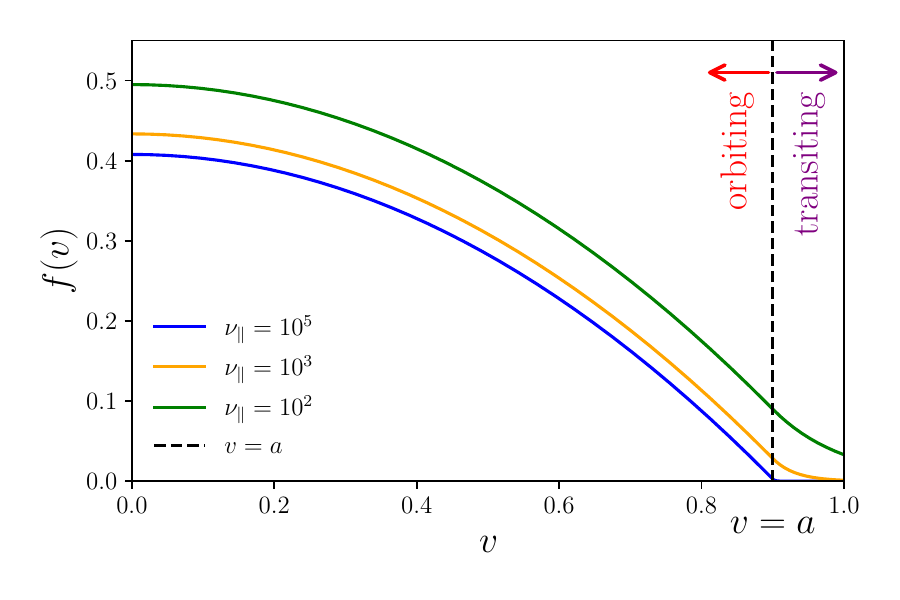}
    \caption{Analytical solutions to equation \eqref{eq: 1D toy model equation dfdt} showing the distribution function profile for parameters $a = 0.9$, $S= 1.0$, $D=1.0$ with varying parallel streaming rates $\nu_\parallel$.}
    \label{fig:1D_toy_model_diffusion_sink_varying_nu_par}
\end{figure}

The equilibrium solution to equation \eqref{eq: 1D toy model equation dfdt} is shown in figure \ref{fig:1D_toy_model_diffusion_sink_varying_nu_par}, demonstrating the importance of resolving the fast advection in this problem. A slow depletion rate in the transit region (larger $\nu_\parallel$) results in enhanced confinement, increasing $f$ in the orbiting region. Likewise, setting $f=0$ in the transit region, as is done with first-order bounce averaging (smaller $\nu_\parallel$), decreases $f$ in the orbiting region, so it is important to get the collisional decay correct in the transit region. Work under the framework of the analytic bounce-average treats the transit region as $f=0$, which is shown not to completely capture the dynamics \citep{bendaniel1962scattering, marx1970effects, Pastukhov_1974, matsuda1986relativistic, harvey1992cql3d, li2025approximating, Rosen_Sengupta_Ochs_Parra_Hammett_2025}. An alternative to the POA algorithm is to solve the OAP equations without the FDP or $H_{\rm orbit}$. Although slowing down the bouncing motion of equation \eqref{eq: Full vlasov equation} by putting the factor $\alpha$ in front of the Poisson bracket would be straightforward, such a modification underestimates the essential sink rate to the problem. Furthermore, a multirate method which incorrectly identifies the fast/slow split of the dynamics as being between purely the slow collisional and fast advection terms in equation \eqref{eq: Full vlasov equation} would advance the collision operator everywhere in a big time step, resulting in enhanced diffusion into the transit region and an enlarged density. The equilibrium solution to equation \eqref{eq: 1D toy model equation dfdt} demonstrates why slowing down the advection or advancing the collision operator in a multi-rate method everywhere is a flawed idea, motivating the POA algorithm, which models both regions to achieve consistent dynamics.

Although this model is motivated by electrostatically confined particles, it also applies to electrostatically lost particles via symmetry. Simulations of magnetic mirrors feature both electrostatically lost ions and electrostatically confined electrons, so it is important that the POA technique works for both \citep{Pastukhov_1974, Najmabadi_Conn_Cohen_1984, Rosen_Sengupta_Ochs_Parra_Hammett_2025}. In the 1D model, one simply has to reverse the order of the terms $v-a$ to $a-v$, and then it's an electrostatically lost region, admitting a similar analytic solution with the infinite nature of $v>0$ handled using the Wiener-Hopf method.

\subsection{Numerical Results} \label{sec: 1D model numerical results}
While equation \eqref{eq: 1D toy model equation dfdt} accepts a straightforward steady-state solution, it is a valuable exercise to solve this numerically as an initial value problem. The POA algorithm is compared to a direct Runge-Kutta 4th-order (RK4) scheme to understand the trade-offs in this method. In the POA integrator, the RK4 method is also used for the time steps in both the full dynamics phase (FDP) and the orbit-averaged dynamics phase (OAP), but the size of the time step is modulated accordingly. In the OAP, the algorithm solves the orbit-averaged equation:
\begin{equation}
\frac{\partial f(v,t)}{\partial t} = H(a-v)\left(D\frac{\partial^2 f}{\partial v^2} + S \right), \label{eq: 1D toy model modulated equation}
\end{equation}
which freezes the evolution $v>a$ while preserving the dynamics of the source and diffusion balance.

\begin{figure}[t]
    \centering
    \includegraphics[width=\linewidth]{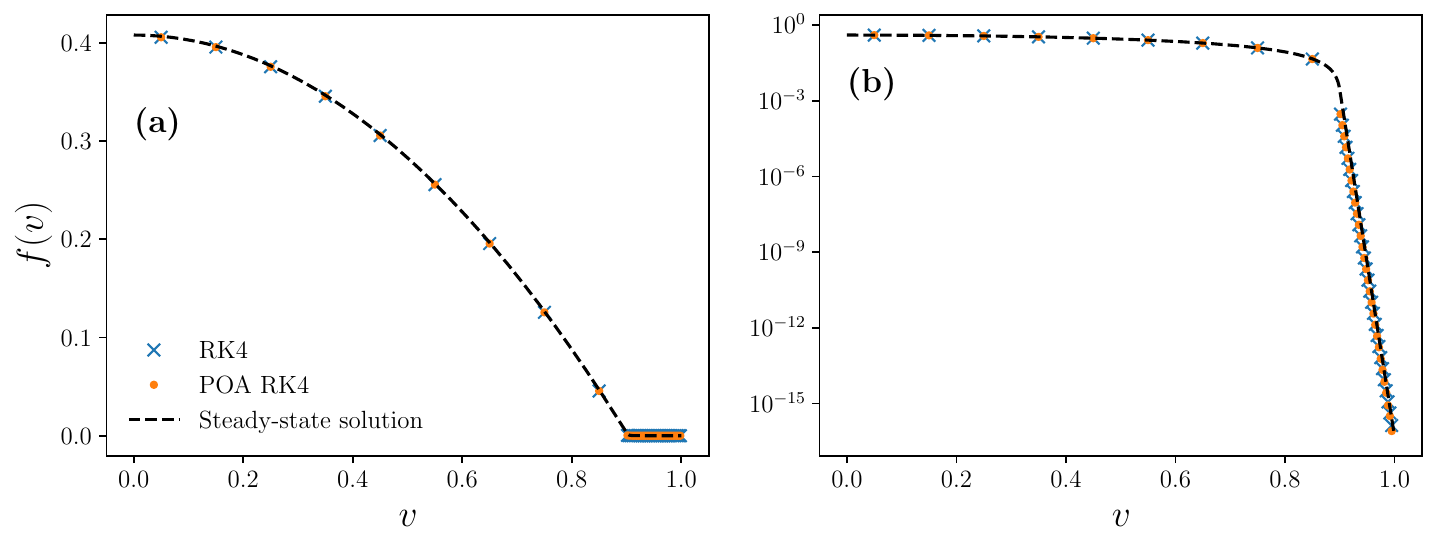}
    \caption{Comparison of steady-state solutions obtained using the POA algorithm (orange dots) versus a direct RK4 method (blue exes) in linear (a) and logarithmic (b) scales. Both simulations evolved to $t=20 L_v^2/D$. The analytic solution (black dashed line) is provided for reference. Parameters $D = 1$, $\nu_\parallel = 10^5$, $a = 0.9$, $L_v = 1.0$, $S = 1$, $\alpha=2D/(\nu_\parallel(\Delta v_{\rm center} )^2) = 2\times 10^{-3}$, $\tau_{\rm FDP} \nu_\parallel= 5$, $\tau_{\rm OAP} D / L_v^2  = 5$, $\Delta v_{\rm orbit} = 1/10$, and $\Delta v_{\rm transit}=1/300$.}
    \label{fig:1D_toy_model_steady_state_comparison}
\end{figure}

For the numerical implementation, the problem is discretized using the finite volume method on a non-uniform velocity mesh $v \in [0, 1]$. The transit region requires higher resolution due to the solution's exponential decay to achieve comparable accuracy in both regions. The orbit region has cell spacing $\Delta v_{\rm orbit}$ and the transit region has cell spacing $\Delta v _{\rm transit}$ with $\Delta v_{\rm orbit} > \Delta v_{\rm transit}$ and domain length in velocity space $L_v=1$. The diffusion operator is implemented using a central difference. Initial conditions $f(v,0) = 0$ for all $v$. The boundary conditions are $f'(0,t) = 0$ (reflecting boundary) and $f(1,t) = 0$ (absorbing boundary). 

The Courant condition governs the time step and numerical stability. The RK4 method is used for all time steps, so a time step near the stability boundary is chosen. For this problem, the factor in front of $\Omega$, the largest complex frequency in the system, is empirically found at $\Omega \Delta t=1.0$, as a larger time step fails to capture the correct diffusive dynamics. The associated Courant condition of this problem is determined by the maximum rate of either the diffusion term or the sink.
\begin{equation}
\Delta t_{\rm FDP} = \left[\frac{2D}{(\Delta v_{\rm transit})^2} + \nu_\parallel\right]^{-1}.
\label{eq: CFL condition}
\end{equation}
On the other hand, the OAP scales $\nu_\parallel$ by the factor $\alpha$, the minimum resolution is increased to be $\Delta v_{\rm orbit}$, and the Courant condition is modified to be
\begin{equation}
\Delta t_{\rm OAP} = \left[\frac{2D}{(\Delta v_{\rm orbit})^2} + \alpha \nu_\parallel\right]^{-1}.
\label{eq: CFL condition OAP}
\end{equation}

The numerical simulations presented employ the following parameters: $D = 1$, $\nu_\parallel = 10^5$, $a = 0.9$, $S = 1$, $\Delta v_{\rm orbit} = 1/10$, and $\Delta v_{\rm transit}=1/300$. For the POA integrator, during the OAP, $\alpha=2D/(\nu_\parallel(\Delta v_{\rm orbit} )^2) = 2\times 10^{-3}$ is used to balance the diffusion term with the advective term. In this scheme, $\tau_{\rm FDP} \nu_\parallel = 5$ and $\tau_{\rm OAP} D /L_v^2 = 5$. The OAP starts the POA integrator, followed by the FDP, and the cycle repeats until each simulation reaches 20 $ L_v^2/D$. In this parameter regime, the POA scheme requires $700$ times fewer steps to reach 20 $ L_v^2/D$ than the purely RK4 method. The formula $(\tau_{ \rm FDP} + \tau_{\rm OAP} )/ (\tau_{\rm FDP} + \alpha \tau_{\rm OAP})$ gives a speed up of $500\times$, which is a close estimate of the correct order of magnitude. The discrepancy here arises because the theoretical formula does not account for the additional restriction due to the velocity resolution, as well as the inverse-sum operation to find the maximum rate for the time-step, which results in a time step smaller than the true rate of each term (this is a common obstacle in high fidelity codes \citep{Francisquez_Cagas_Shukla_Juno_Hammett_2025, mandell2024gx}).

\begin{figure}[t]
    \centering
    \includegraphics[width=\linewidth]{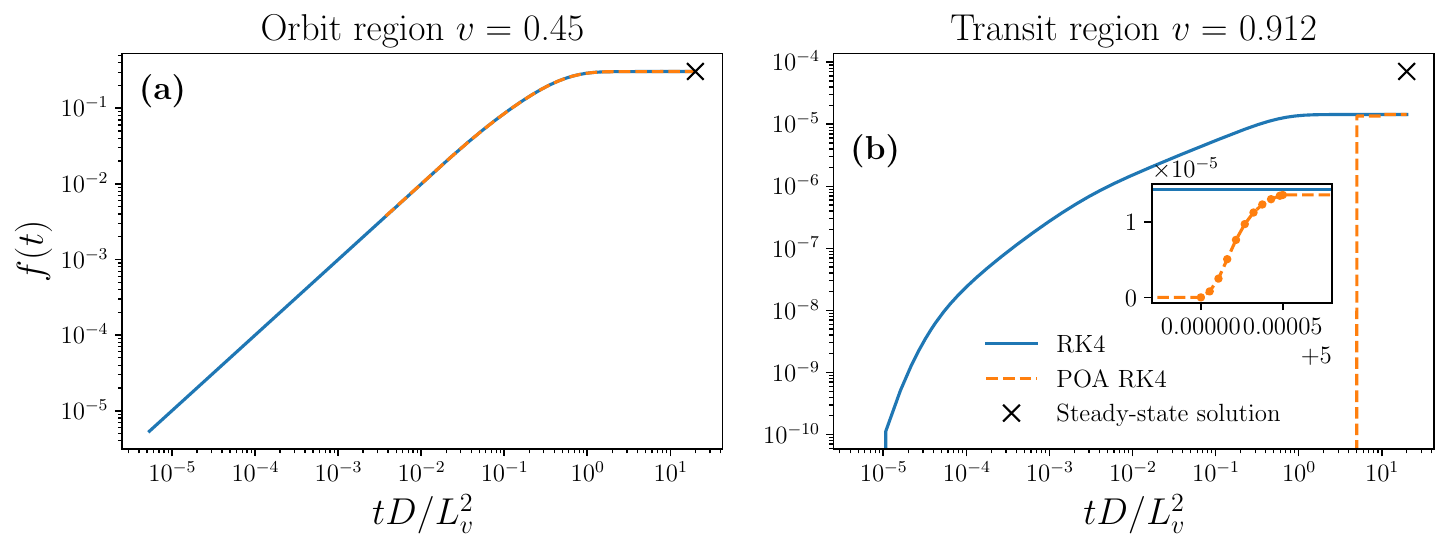}
    \caption{Time evolution of the distribution function using the RK4 (solid blue) and the POA algorithm (dashed orange) inside (a: $v = 0.45$) and outside (b: $v = 0.912$) the orbit/transit separatrix. Parameters are identical to figure \ref{fig:1D_toy_model_steady_state_comparison}. The inset shows the first FDP after a long OAP. The inset is shown in a linear x-y scale where the time axis starts at $\tau_{\rm OAP} - 0.6\tau_{\rm FDP}$ and extends to $\tau_{\rm OAP} + 1.6\tau_{\rm FDP}$. The POA RK4 line in the inset is shown with dots on each time step to show the rapidness of evolution during the FDP.}
    \label{fig:1D_toy_model_short_time_behavior}
\end{figure}

Figure~\ref{fig:1D_toy_model_steady_state_comparison} demonstrates that both the POA algorithm and direct RK4 integration converge to an equilibrium that agrees with each other. The exponential decay in the transit region is accurately captured. Both numerical methods agree, indicating that the POA multirate method introduces no numerical artifacts.
Figure~\ref{fig:1D_toy_model_short_time_behavior} reveals essential differences in the transient evolution between the two approaches, normalized to a diffusion time. Within the orbiting region at $v = 0.45$, the two integrators agree exactly. This is expected because the problem is unmodified in this region during the OAP. However, in the transit region at $v = 0.912$, the two integrators differ significantly. In the first OAP, the initial condition is set to zero, resulting in no points on the log scale. The first FDP starts at $t D / L_v^2 = 5$ and quickly diffuses $f$ into the transit region, requiring only a short amount of time to reach steady-state. The values of $f$ in the transit region are held fixed for the next OAP. After a few cycles of the POA scheme, the steady-state solution is reached and agrees with the RK4 integrator.
\section{Five-Equation ODE Model}
\label{sec: five equation toy model}

To further demonstrate and test the POA algorithm, a model based on 5 coupled ordinary differential equations (ODEs) is developed to capture the fast advection and slow collisions in a magnetic mirror configuration. 
The model is motivated by considering two trajectory contours, depicted in figure \ref{fig: schematic of bouncing points}, where one trajectory is periodic, representing an orbiting contour, and the other is free-streaming, with collisional scattering (diffusion) between them. The model will evolve $f$ at three points along the orbit contour and at two points along the transit contour.
The coefficients in the 5-equation ODE model can be varied to consider test problems with various types of features.

The 5-equation ODE model we present here can be thought of as a very coarse discretization of
the model PDE in equation \eqref{eq:toy-OA}. 
Consider a uniform grid in $\theta$, and define $f_{o,j}(t) = f(\theta_j, v_o,t)$ and $f_{t,j}(t) = f(\theta_j,v_t,t)$, where $v_o$ labels a trajectory in the orbit region and $v_t$ labels a transit trajectory.
First, focus on the fast advection term $\Omega \partial f/\partial \theta$ in the orbit region. 
We use centered differencing in the orbit region to avoid any numerical dissipation,
$\Omega \partial f / \partial \theta \approx \Omega (f_{o,j+1}-f_{o,j-1})/(2 \Delta \theta) = \nu_{\parallel}(f_{o,j+1}-f_{o,j-1})$, where $\nu_\parallel = \Omega / (2 \Delta \theta)$.
Using periodicity in the index $j$ (from periodicity in $\theta$), we see that we need at least 3 points to have a non-trivial result (this operator vanishes if there are only 2 points). 

In the transit region, the parallel advection operator causes particles to be rapidly lost. To handle the boundary conditions easily, we use first-order upwind finite differencing in the transit region, $\Omega \partial f/\partial \theta \approx \Omega (f_{t,j} - f_{t,j-1}) / \Delta \theta$.
For simplicity, we use just two points in the transit region. With zero influx boundary conditions, $f_{t,0} = 0$, we get the $\nu_\parallel$ terms in equations (\ref{eq: first statement of five equation model eq 4}-\ref{eq: first statement of five equation model eq 5}). (For simplicity, we use the same value of $\nu_\parallel$ for the transit points and orbit points.)
One could consider a simpler model with only 1 point in the transit region, but we found it helpful to consider the additional downstream point $f_{t2}$ in the exhaust region, which is fed indirectly via advection from $f_{t1}$.

The collisional diffusion term in velocity space in equation \ref{eq:toy-OA} is simplified by representing it as a BGK-type operator that damps towards a solution that is uniform between the orbit points $f_{oj}$ and adjacent transiting points, $-\nu_{cj}(f_{oj}-f_{tj})$. 
Particles in $f_{o1}$ could diffuse to $f_{t1}$, and particles in $f_{o2}$ and $f_{o3}$ could diffuse onto a transit trajectory in the bottom half of the figure, but for simplicity we use $f_{t1}$ as representative of $f$ in both the upper and lower transit regions, so the collisional damping term for $f_{oj}$ is $-\nu_{cj}(f_{oj}-f_{t1})$.  This gives to the $\nu_{cj}$ terms in equations (\ref{eq: first statement of five equation model eq 1}-\ref{eq: first statement of five equation model eq 3}) below.  
The collisional loss rate $\nu_{cj}$ is allowed to vary to consider such cases.
To ensure the BGK-type term by itself conserves particles, the $\nu_{cj}$ losses from each of the $f_{oj}$ are added as source terms for $f_{t1}$ in equation (\ref{eq: first statement of five equation model eq 4}).  
Thus the BGK-type terms by themselves (setting $\nu_\parallel=S_j =0$ in equations(\ref{eq: first statement of five equation model eq 1}-\ref{eq: first statement of five equation model eq 5})) damp towards equipartition, $f_{o1} = f_{o2} = f_{o3} = f_{t1}$.

Putting this all together, the 5-equation ODE model is:
\begin{figure}[t]
\centering
\includegraphics[width=\textwidth]{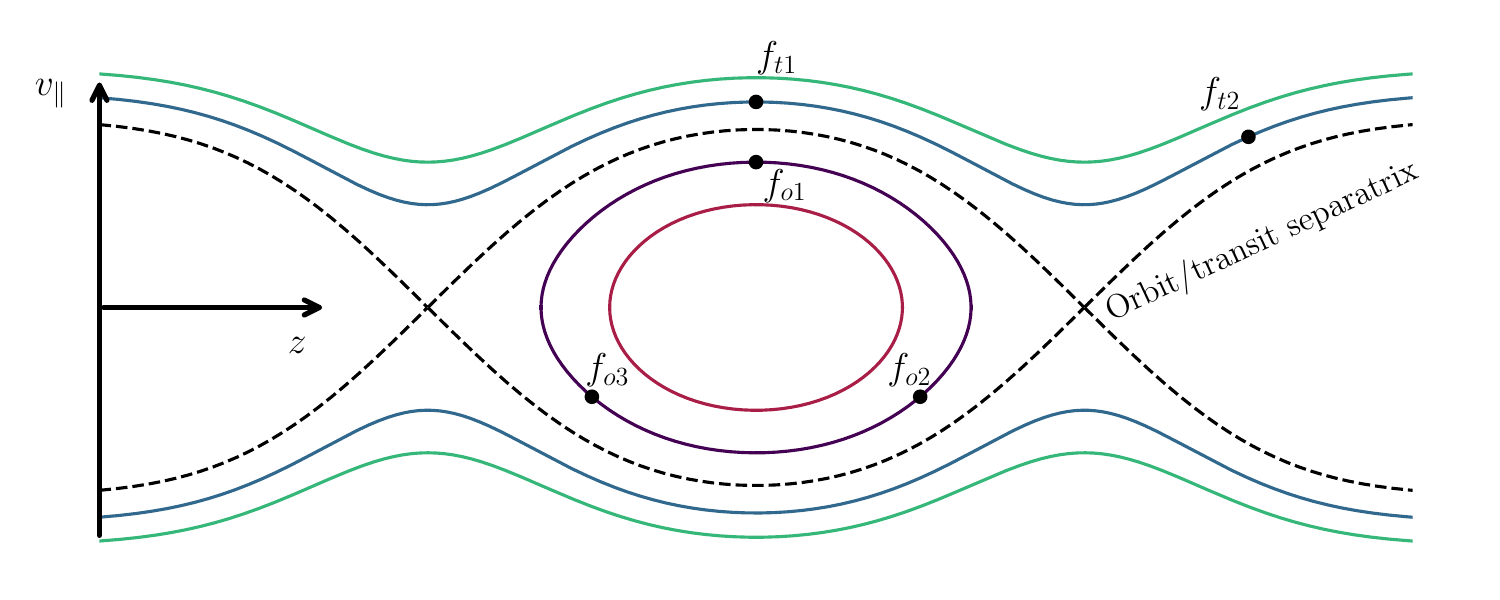}
\caption{Schematic representation of the phase space diagram in a magnetic mirror. $f_{o1}$, $f_{o2}$, and $f_{o3}$ are orbiting regions, while $f_{t1}$ and $f_{t2}$ are transiting with an exhaust.}
\label{fig: schematic of bouncing points}
\end{figure}
\begin{align}
    \frac{d f_{o1}}{d t} &= - \nu_\parallel (f_{o2}-f_{o3}) + \nu_{c1}(f_{t1}-f_{o1}) + S_1, \label{eq: first statement of five equation model eq 1}\\
    \frac{d f_{o2}}{d t} &= - \nu_\parallel (f_{o3}-f_{o1}) + \nu_{c2}(f_{t1}-f_{o2}) + S_2 ,\\
    \frac{d f_{o3}}{d t} &= - \nu_\parallel (f_{o1}-f_{o2}) + \nu_{c3}(f_{t1}-f_{o3}) + S_3 ,
    \label{eq: first statement of five equation model eq 3}\\
    \frac{d f_{t1}}{d t} &= - \nu_\parallel f_{t1}  +  \nu_{c1}(f_{o1}-f_{t1}) + \nu_{c2}(f_{o2}-f_{t1}) + \nu_{c3}(f_{o3}-f_{t1}) ,
    \label{eq: first statement of five equation model eq 4}\\
    \frac{d f_{t2}}{d t} &= - \nu_\parallel (f_{t2} - f_{t1}) \label{eq: first statement of five equation model eq 5}.
\end{align}
The model incorporates two characteristic rates that create the multiscale structure: $\nu_{ci}$ is the slow collisional coupling between the orbit and transit region; $\nu_\parallel$ is the fast advection rate describing rapid orbital motion in the orbit region and the fast exhaust rate in the transit region. 
This set of five coupled ordinary differential equations (ODEs) is reminiscent of the five-equation ODE model used in \citet{orszag1973statistical} to study renormalized turbulence theories.

The POA algorithm is applied to the five-equation model as described in section \ref{sec:POA-algorithm}.
During the OAP, the transit region is frozen (by setting $H_{\rm orbit} = 0$ in the following equations), and the orbital advection is slowed by the factor $\alpha$, by solving the modified equations:
\begin{align}
    \frac{d f_{o1}}{d t} &= - \alpha \nu_\parallel (f_{o2}-f_{o3}) + \nu_{c1}(f_{t1}-f_{o1}) + S_1, 
    \label{eq: first statement of five equation model eq 1 orbit average phase}
    \\
    \frac{d f_{o2}}{d t} &= - \alpha \nu_\parallel (f_{o3}-f_{o1}) + \nu_{c2}(f_{t1}-f_{o2}) + S_2,\\
    \frac{d f_{o3}}{d t} &= - \alpha \nu_\parallel (f_{o1}-f_{o2}) + \nu_{c3}(f_{t1}-f_{o3}) + S_3,
    \label{eq: first statement of five equation model eq 3 orbit average phase}
    \\
    \frac{d f_{t1}}{d t} &= H_{\rm orbit} ( - \alpha \nu_\parallel f_{t1}  +  \nu_{c1}(f_{o1}-f_{t1}) + \nu_{c2}(f_{o2}-f_{t1}) + \nu_{c3}(f_{o3}-f_{t1}) ) , 
    \label{eq: first statement of five equation model eq 4 orbit average phase}
    \\
    \frac{d f_{t2}}{d t} &= H_{\rm orbit} ( - \alpha \nu_\parallel (f_{t2} - f_{t1})) 
    \label{eq: first statement of five equation model eq 5 orbit average phase}.
\end{align}
The FDP solves these same equations with $\alpha=1$ and $H_{\rm orbit}=1$.

Let us consider some of the general properties of this 5-equation ODE model (in the $\alpha=1$ case, but one can generalize by replacing $\nu_\parallel \rightarrow \alpha\nu_\parallel$).
In the limit where collisions are neglected ($\nu_{ci} = 0$), and there is no source ($S = 0$), the system reduces to pure fast advection among the orbiting points:
\begin{align}
\frac{d f_{o1}}{d t} &= - \nu_\parallel (f_{o2} - f_{o3}), \\
\frac{d f_{o2}}{d t} &= - \nu_\parallel (f_{o3} - f_{o1}), \\
\frac{d f_{o3}}{d t} &= - \nu_\parallel (f_{o1} - f_{o2}).
\end{align}
This subsystem conserves both number density, $f_{o1} + f_{o2} + f_{o3}$, and a wave energy-like quantity, $f_{o1}^2 + f_{o2}^2 + f_{o3}^2$. The normal mode frequencies are:
\begin{align}
\omega_{0} &= 0, \\
\omega_{\parallel \pm} &= \pm  \sqrt{3} \nu_{\parallel},
\end{align}
and oscillations in time occur with a period $T = 2 \pi / \omega_\parallel$. 
When collisions are included, there will be some damping.  In the uniform damping limit $\nu_{ci} = \nu_c$, then these eigenvalues are all shifted by $-i \nu_c$.
When the full system includes collisions, particle conservation is maintained: the total $f_{o1} + f_{o2} + f_{o3} + f_{t1} + f_{t2}$ is conserved when the source $S$ and final exhaust term $\nu_\parallel f_{t2}$ are excluded.

When collisions are neglected, the dynamics in the transit region for $f_{t1}$ and $f_{t2}$ are decoupled from the solution for $f_{o1}$, $f_{o2}$, and $f_{o3}$ in the orbit region. The solution in the transit region corresponds to just strongly decaying dynamics, with 2 large damped eigenfrequencies, both with $\omega = - i \nu_\parallel$, representing the rapid loss of particles in the transit region.

For the full system of equations,
equations (\ref{eq: first statement of five equation model eq 1 orbit average phase}-\ref{eq: first statement of five equation model eq 5 orbit average phase}),
summing the first 3 equations together in equilibrium shows that the net flux into the transit region is
$\nu_{c1} (f_{o1} - f_{t1}) 
+ \nu_{c2} (f_{o2} - f_{t1}) 
+ \nu_{c3} (f_{o3} - f_{t1}) 
= S_1 + S_2 + S_3$, which depends only on the total source and is independent of the choice of $\nu_{ci}$.  
Thus, the equilibrium solution in the transit region is 
$f_{t1} = f_{t2} = (S_1 + S_2 + S_3) / \nu_\parallel$.
This is why the POA algorithm turns off the evolution of $f$ in the transit region during the OAP when $\alpha \ll 1$, because $f$ in the transit region would be too big if $\nu_\parallel$ is slowed by down to $\alpha \nu_\parallel$ with $\alpha \ll 1$.

In the parameter regime of high orbit frequencies that we are interested in, $\nu_\parallel \gg \nu_{cj}$, the lowest order orbit-averaged equilibrium solution is just $f_{o1} = f_{o2} = f_{o3} = f_o = (S_1+S_2 + S_3) / (\nu_{c1} + \nu_{c2} + \nu_{c3})$, so the lowest order solution in the orbiting region is independent of $\nu_\parallel$.
The solution in the transit region is smaller than in the orbiting region by a factor of $f _{t1} / f_o = (\nu_{c1}+\nu_{c2}+\nu_{c3})/\nu_\parallel$ and is neglected in lowest order orbit-averaged theory.

The fact that the equilibrium solution in the orbit region is independent of $\nu_\parallel$  is the motivation for introducing the $\alpha$ parameter into these equations, to slow down the frequency in the orbiting region and allow larger time steps to get to equilibrium more quickly.


To determine the time step, a model Courant condition for an RK4 scheme is constructed by analyzing the problem's rates. 
The time step $\Delta t$ should be chosen so that $\omega \Delta t$ lies within the stability boundary for RK4 for all complex eigenfrequencies $\omega = \omega_r + i \omega_i$ of the system.  We approximate this conservatively by $(|\omega_r| + |\omega_i|) \Delta t < 2.3$, where $\omega_r$ is the maximum frequency of the Hamiltonian terms and $\omega_i$ is the maximum damping rate from the dissipative terms, so the time step is:
\begin{equation}
\Delta t  = \frac{2.3}{\max(\nu_{ci}) + \alpha \sqrt{3} \nu_\parallel}.
\label{eq:RK4-dt}
\end{equation}
The coefficient of $2.3$ is chosen as it lies $\approx 20\%$ away from the stability boundary for RK4. Operating close to the stability boundary of RK4 introduces numerical damping, which is important to be mindful of. The maximum number of steps in $\Delta t$ is set so that the FDP and OAP last a certain number of time steps, rather than taking a small last step to end on the exact time for that phase.
During the FDP ($\alpha = 1)$, the Courant condition is unaffected compared to the full problem. During the OAP, $\alpha \ll 1$ can be chosen to increase the time step.   Bounce-averaging is formally still valid as long as $\max(\nu_{ci}) / (\alpha \nu_\parallel) \ll 1$.
We will typically choose $\alpha$ at the edge of this limit, with $\alpha = \max(\nu_{ci}) / (\sqrt{3} \nu_\parallel)$ so that dissipative and oscillatory terms have comparable rates.

This works well for the case of uniform sources ($S_i = S$) and uniform damping ($\nu_{ci} = \nu_c$) but in cases with strong non-uniformities in the sources or in the damping, this can cause errors of order $\max(\nu_{ci}) / (\alpha \nu_\parallel) \sim \epsilon / \alpha$, where $\epsilon = \max(\nu_{ci})/\nu_\parallel$ is the orbit-averaging expansion parameter.
Below, we will discuss some strategies for dealing with these cases, but one approach is to choose an intermediate $\alpha$. For example, there are cases of practical interest where $\epsilon \sim 10^{-5}$, so one could choose $\alpha = 10^{-3}$ to get a speedup of a factor of $1/\alpha \sim 10^3$ with an error of order $\epsilon/\alpha \sim$ 1\%.

In this section, four conditions are explored to study the optimal approach for running the POA scheme for different types of systems: localized ($S_1=3 S$, $S_2 = S_3 = 0$) versus distributed ($S_i = S$) source, and localized ($\nu_{ci}=\nu_{cj}=0, \nu_{ck}=3\nu_c)$ versus distributed ($\nu_{ci}=\nu_c$) orbit/transit coupling.

\subsection{Distributed Source and Orbit/Transit Coupling}
\label{sec: subsection distributed source and sink}

\subsubsection{Problem Setup}
\label{sec: subsubsection distributed source and sink problem setup}

The five-equation model with a distributed source and orbit/transit coupling are represented by equations \eqref{eq: first statement of five equation model eq 1 orbit average phase} - \eqref{eq: first statement of five equation model eq 5 orbit average phase} when $S_1=S_2=S_3=S$ and $\nu_{c1}=\nu_{c2}=\nu_{c3}=\nu_c$. During the FDP, $\alpha=H_{\rm orbit} = 1$ and during the OAP, $H_{\rm orbit}=0$ and $\alpha$ is specified below. The FDP steady-state solution can be found analytically:
\begin{align}
    f_{o1,\infty} = f_{o2,\infty} = f_{o3,\infty} &= \frac{S}{\nu_c}\left(1 + 3 \varepsilon\right), \label{eq: analytic steady state solution ft distributed source, distributed sink}\\
    f_{t1,\infty} = f_{t2,\infty} &= \frac{3S}{\nu_c}\varepsilon . \label{eq: analytic steady state solution expander loss distributed source, distributed sink}
\end{align}
where $\varepsilon = \nu_c / \nu_\parallel$. After converging in the transiting region, the steady state in the orbiting region during the OAP is the same as for the FDP because the factor of $\alpha$ can be divided out during the mathematical derivation.
\begin{figure}[t]
    \centering
    \includegraphics[width=\linewidth]{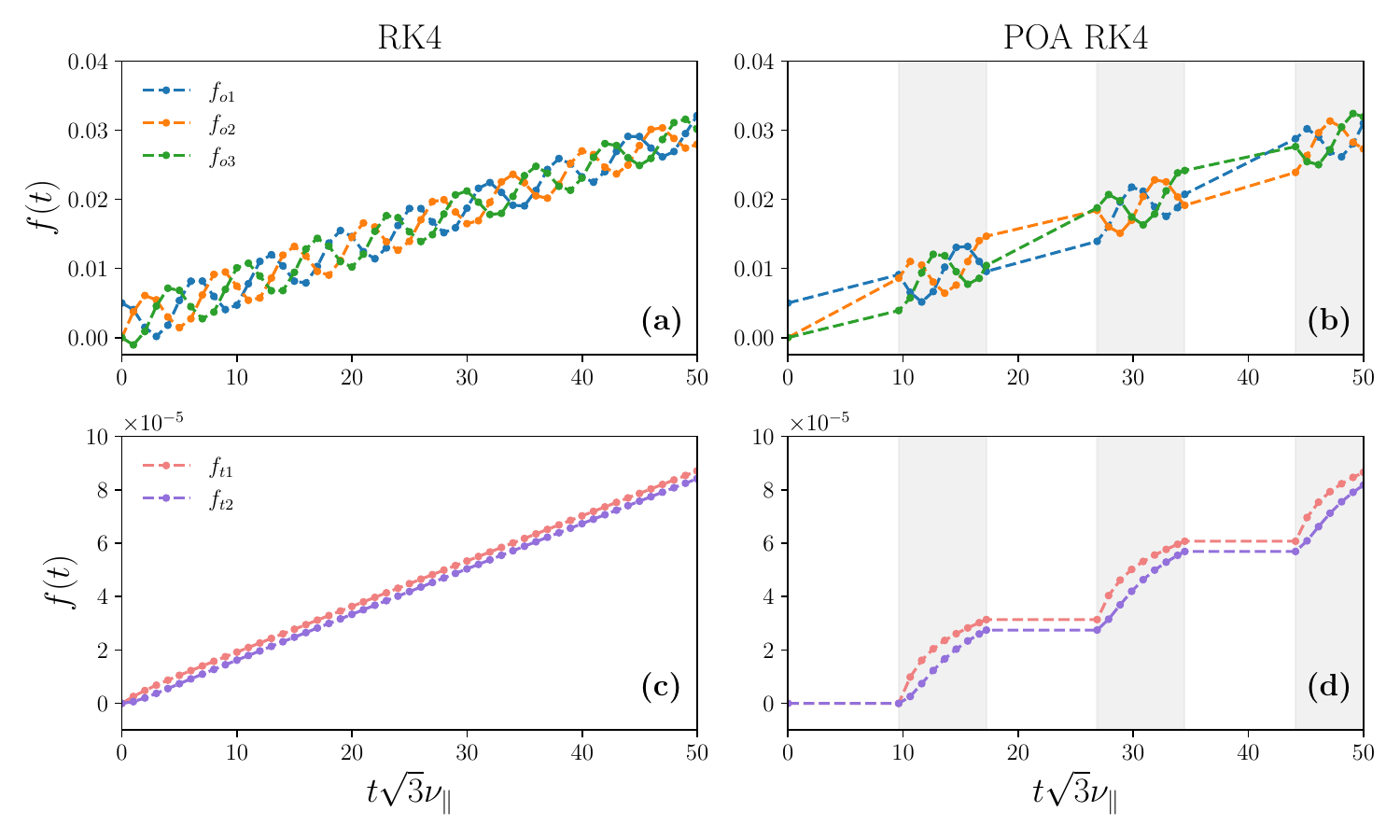}
    \caption{Short-time comparison between standard RK4 integration (left) and the POA algorithm (right), in the orbiting (a and b) and transiting (c and d) regions of $f$. Grey regions show the FDP, while the OAP has a white background. Parameters: $S = 1$, $\nu_c = 1$, $\nu_{\parallel} = 10^3$, $\alpha = 0.1$, $\tau_{\rm FDP} \nu_\parallel=1.4 \pi $, $\tau_{\rm OAP} \nu_c= 1/180$, distributed source, distributed coupling.}
    \label{fig: fiveeq beginning distributed source and loss}
\end{figure}

\subsubsection{Numerical Results}

The model described in section \ref{sec: subsubsection distributed source and sink problem setup} is implemented in Python and accelerated with JAX in double precision. Particular results of interest include the evolution of transients and long-term steady-state behavior, as well as the exploration of extreme use cases of the algorithm.

Figure~\ref{fig: fiveeq beginning distributed source and loss} demonstrates the algorithm's behavior in an initial transient, using parameters chosen to provide a clear visualization of the multiscale dynamics: $S = 1$, $\nu_c = 1$, $\nu_{\parallel} = 10^3$, and $\alpha = 0.1$. The initial conditions have all values set to zero, except $f_{o1} = 0.005$ to initiate advective scale oscillations. The POA evolves the FDP for $\tau_{\rm FDP} \nu_\parallel=1.4 \pi $ and the OAP is evolved for $\tau_{\rm OAP} \nu_c = 1/180$, which only requires a single large time step. Although the coefficient in front of the RK4 time-step is set to 2.3 in equation \eqref{eq:RK4-dt} for most sections, the large coefficient induces numerical damping, so here the coefficient is set to 1.0 to illustrate the oscillations in $f_o$. 
In the regular RK4 solution, $f_{o1}$, $f_{o2}$, and $f_{o3}$ oscillate with a period of $T = 2 \pi / \sqrt{3}\nu_\parallel$, as expected. $f_{t1}$ and $f_{t2}$ grow monotonically and slowly, supporting the use of the POA multiscale time-stepper. The POA algorithm preserves the amplitude of the oscillations between $f_{o1}$, $f_{o2}$, and $f_{o3}$ between OAP-FDP phases. There is minimal difference between $f_{t1}$ and $f_{t2}$  in the RK4 and POA RK4 schemes, so both systems end around the same value. Here, the RK4 integrator required 51 evaluations, while the POA scheme required 25, representing only a 2x speed up, but this is a case where we are using small time steps in the OAP phase to look at the short time transient dynamics, so $\alpha=0.1$ was not as small as it could be for getting to steady-state, which we consider next.

\begin{figure}[t]
    \centering
    \includegraphics[width=\linewidth]{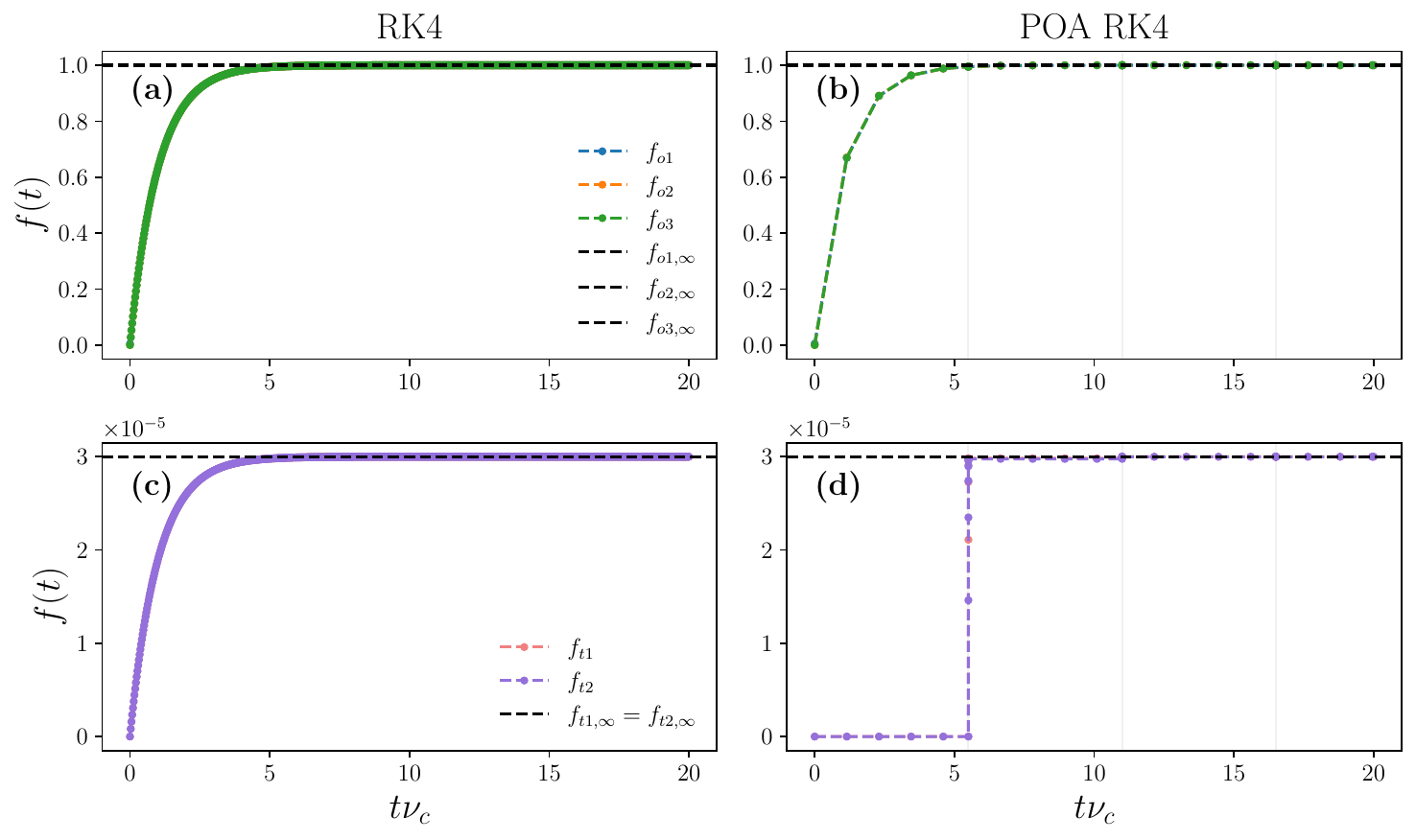}
    \caption{Long-term steady-state behavior comparing standard RK4 (left) and the POA algorithm (right), in the orbiting (a and b) and transiting (c and d) regions of $f$. Grey regions show the FDP, while the OAP has a white background. Parameters: $S = 1$, $\nu_c = 1$, $\nu_{\parallel} = 10^5$, $\alpha = \nu_c / (\sqrt{3} \nu_\parallel)$, $\tau_{\rm FDP} \nu_\parallel = 4.7 \sqrt{3}$, $\tau_{ \rm OAP}\nu_c = 5.5$, distributed source, distributed coupling, and simulating to $t = 20 \nu_c^{-1}$. Black dashed lines show the analytical steady-state solutions. For fast rendering, only every 2,000th point in the RK4 method is plotted.}
    \label{fig: 5 equation model long term behavior distributed source and loss}
\end{figure}

For long-term evolution, figure \ref{fig: 5 equation model long term behavior distributed source and loss} examines a more extreme parameter regime: $S = 1$, $\nu_c = 1$, $\nu_{\parallel} = 10^5$, and $\alpha = \nu_c/(\sqrt{3}\nu_\parallel)$. $\nu_{\parallel} = 10^5$ is chosen because it is closest to the parameter regime that \citet{Francisquez_Rosen_Mandell_Hakim_Forest_Hammett_2023} investigated. Here, the POA RK4 evolves the FDP for $\tau_{\rm FDP} \nu_\parallel = 4.7 \sqrt{3} $ and the OAP evolves for $\tau_{\rm OAP} \nu_c= 5.5$. 
The error in the final state is taken as the fractional error $|f_{i}(t_{\rm final}) - f_{i,\infty}|/f_{i,\infty}$, where $f_{i}(t_{\rm final})$ is the value of $f_i$ at the last time step and $f_{i,\infty}$ is the analytic steady-state solution for the quantity $f_i$. For the RK4 integrator, all values of $f$ end with fractional errors of $2.06\times 10^{-9}$. In the POA RK4 scheme, the $f_o$ values end with a fractional error of $4.15 \times 10^{-9}$, while $f_{t1}$ ends with an error of $1.35 \times 10^{-7}$ and $f_{t2}$ ends with a fractional error of $2.10\times 10^{-7}$. The POA scheme reaches machine precision levels of error when run for $5\times$ longer. In this case, the RK4 algorithm took $1,506,140$ steps while the POA scheme only took $40$, resulting in a $37,600 \times$ reduction in the number of steps.

\begin{figure}[t]
    \centering
    \includegraphics[width=\linewidth]{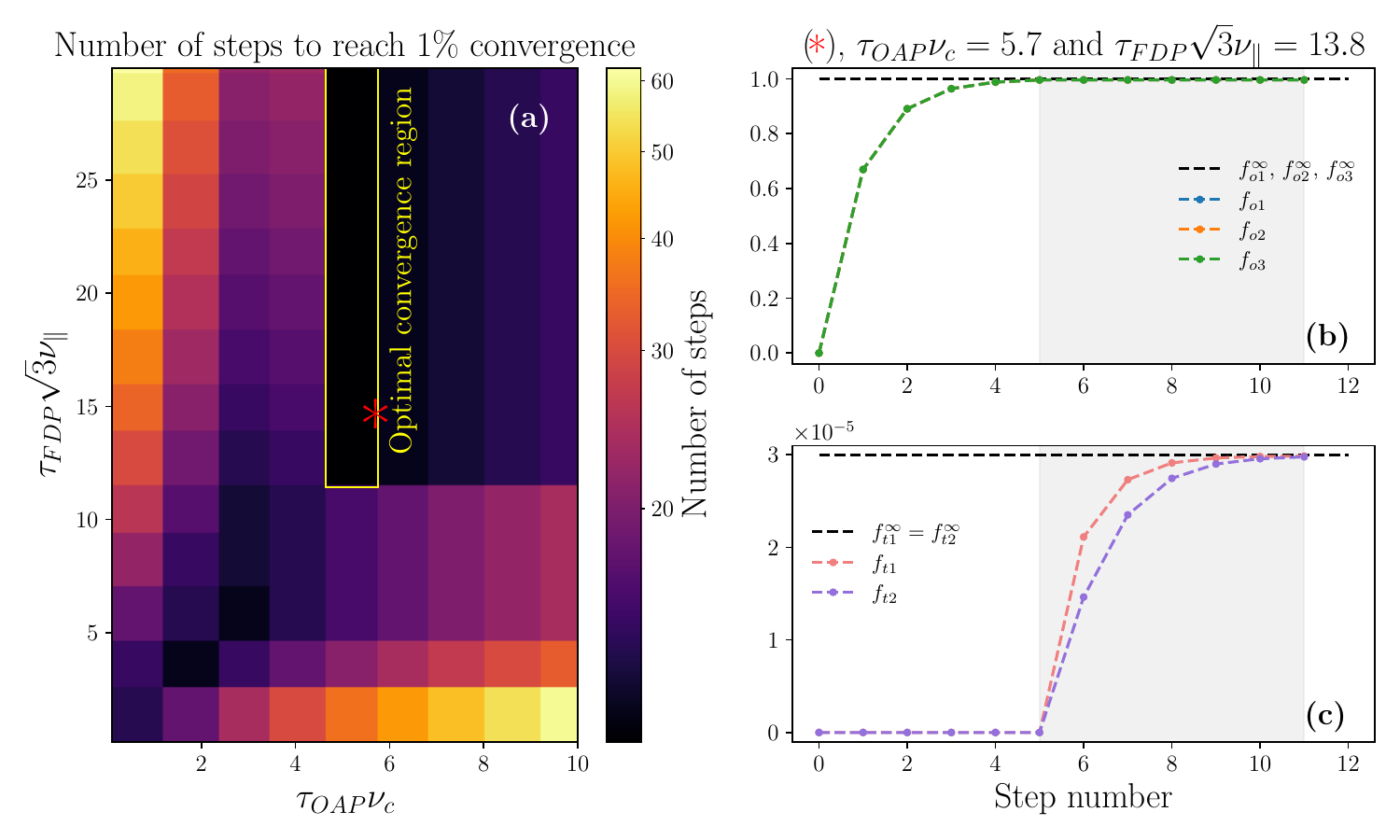}
    \caption{(a) Total number of RK4 steps needed to reach $1\%$ convergence comparing $f_{t2}$ to its analytic value, varying the amount of time spent in the FDP and OAP. Orbiting (b) and transiting (c) solutions are shown in the optimal convergence region, at the red asterisk (a) (11 steps, $31,500 \times$ speedup). Grey regions show the FDP, while the OAP has a white background. Parameters: $S = 1$, $\nu_c = 1$, $\nu_{\parallel} = 10^5$, $\alpha = \nu_c / (\sqrt{3} \nu_\parallel)$, distributed source, distributed coupling.}
    \label{fig: number of steps until fiveq converged}
\end{figure}

To continue, it is essential to determine the optimal time spent in each phase to achieve steady state. The five-equation model is defined as converged when all five variables ($f_{o1}$, $f_{o2}$, $f_{o3}$, $f_{t1}$, $f_{t2}$) are within $1\%$ of their analytic value. To reduce the oscillations in $f_o$, the initial condition is set to zero. The number of steps until convergence is shown in figure \ref{fig: number of steps until fiveq converged}a, with yellow lines bounding the region with the minimum number of steps. The minimum number of steps is 11 for the POA algorithm (5 in the OAP and 6 in the FDP), whereas integrating the whole system with RK4 takes 347,000 steps, representing a $31,500\times$ speedup. 
The results indicate that the optimal approach (though there are others almost as good) for running this algorithm for this case involves using one OAP that evolves for many collision times to reach steady-state in the orbiting region, followed by one FDP to allow $f_{t1}, f_{t2}$ to reach equilibrium as well. Using a longer FDP does not improve convergence here because it takes the same number of steps in the FDP after the OAP to converge, and the simulation ends before the full $\tau_{ \rm FDP}$ is reached. Using a longer $\tau_{\rm OAP}$ compared to the optimum increases the total number of steps during the OAP, while a smaller number of steps would give a similar convergence. 
(A side note: $\tau_{\rm FDP} \sqrt 3 \nu_\parallel = 13.8$ and $\tau_{\rm OAP} \nu_c = 5.7$ correspond to $\tau_{\rm FDP} / T = 2.2$ and $\tau_{\rm OAP} / T = 0.9$, meaning that the system executes 2.25 orbits during the FDP and 0.9 orbits during the OAP.)

A valley exists in a narrow linear region along the diagonal in the bottom left corner of figure \ref{fig: number of steps until fiveq converged}a. 
This indicates that rather than doing a single OAP with 5 steps and a single FDP with 6 steps, one can get almost the same efficiency as the optimal approach by iterating more between phases with each phase lasting a shorter time (as long as each individual step in a phase is with the maximum $\Delta t$ set by equation \eqref{eq:RK4-dt} to achieve the maximum error reduction per step).
That is, one could do 2 steps of an OAP followed by 2 steps of an FDP and repeat, to converge with 12 total steps (6 in OAP and 6 in FDP).
(One might have wondered if the system converges faster by iterating between OAP and FDP more often, but these results show that the net efficiency is about the same.)
Because $f_{ti}$ in the transit region is smaller than $f_{oi}$ in the orbit region by a factor of the orbit-averaging expansion parameter $\varepsilon = \nu_c / \nu_\parallel = 10^{-5} \ll 1$, the solution in the orbit region is very weakly dependent on the solution in the transit region (though the solution in the transit region depends strongly on the solution in the orbit region.).
Thus one can use a single OAP to advance $f_{oi}$ in the orbit region to convergence (as long as the error tolerance is larger than $\varepsilon$, which is satisfied in this example where the error tolerance is 1\%), and then use a single FDP to advance the transit region to convergence as well.

\subsubsection{Discussion}

The distributed-source, distributed-orbit/transit-coupling case provides the cleanest demonstration of the POA algorithm. When the source and damping (orbit/transit-coupling) terms are uniform around the orbits (i.e., they are already equivalent to their orbit-average), then the OAP and FDP describe the same steady-state physics, and the analytic steady state equations \eqref{eq: analytic steady state solution ft distributed source, distributed sink}–\eqref{eq: analytic steady state solution expander loss distributed source, distributed sink} are recovered by both the direct RK4 and the POA integrator to machine precision.

The practical consequence of this is that it allows $\alpha$ to be quite small, and we have chosen it to be $\alpha = \nu_c / (\sqrt{3} \nu_\parallel)$, which makes the oscillatory and damping terms comparable, allowing a very large time step.
The POA algorithm attains very large explicit speedups in this configuration: with the parameters studied the POA scheme reduced explicit RK4 evaluations from $\sim 3.47\times10^{5}$ to 11 steps, consistent with an expected speedup of approximately $\alpha / 3 \sim \nu_\parallel/(3 \nu_c) = 1/(3 \varepsilon)$.

The parameter-scan for convergence here indicate that a general strategy that will work fairly well for some problems: rather than have a fixed number of steps for each phase, one can repeat OAP steps until the orbiting region converges, and then repeat FDP steps until the transiting region converges.
However, in a problem such as the one in \citet{Francisquez_Rosen_Mandell_Hakim_Forest_Hammett_2023}, there can be a non-linear coupling of the solution $f_{t}$ and $f_o$ in the transit and orbit regions (such as through the electrostatic potential in a mirror problem, i.e., the orbiting/transiting boundary can shift as the potential varies).
In this case, it might be necessary to do shorter OAPs and FDPs (so that the potential doesn't change to much in a single phase) and iterate more frequently between them in order for the potential to converge.
The parameter scan in figure \ref{fig: number of steps until fiveq converged} shows this is almost as efficient as using a single long OAP followed by a single long FDP.

Taken together, the distributed source/orbit/transit-coupling results establish a “best case” for the POA integrator. They show the algorithm accurately preserves both fast advective structure and slow collisional evolution while delivering very large computational savings. This makes the distributed–distributed configuration the natural starting point for applying the POA scheme to more complex kinetic codes; deviations from that ideal (localized fuelling, localized orbit/transit coupling, or strong nonlinear coupling) then become the cases where the orbit-averaging corrections, explored in later subsections, are most useful.
\subsection{Localized Source, Distributed Orbit/Transit Coupling}

We now test the POA algorithm on a case where the source is not uniform along orbits but is localized somewhere.
Localized sources are interesting for simulations of magnetic mirrors because many experiments have a beam-like fueling mechanism, such as neutral beam injection \citep{Endrizzi_Anderson_Brown_Egedal_Geiger_Harvey_Ialovega_Kirch_Peterson_Petrov_etal._2023}. Thus, it is worthwhile to explore how the POA scheme behaves depending on the source's localization.\footnote{In some cases, it is not too difficult to find the orbit average of a localized source by expressing it in terms of constants of the motion, or even by numerically averaging the source. But here we consider a localized source.}

To modify the five-equation model for source localization, a parameter $\kappa \in [0,1]$ is introduced such that $\kappa = 0$ is a perfectly orbit distributed source and $\kappa = 1$ is purely localized to $f_{o1}$. The model in equations \eqref{eq: first statement of five equation model eq 1 orbit average phase} - \eqref{eq: first statement of five equation model eq 5 orbit average phase} is modified so that  $S_1=S (1 - \kappa) + 3 S \kappa$ and $S_2=S_3= S (1 - \kappa)$ so that $S_1+S_2+S_3=3S$ and $\nu_{c1}=\nu_{c2}=\nu_{c3}=\nu_c$.
During the FDP, $\alpha=H_{\rm orbit} = 1$ and during the OAP, $H_{\rm orbit}=0$ and $\alpha$ is specified below. 
The steady-state equilibrium in the OAP with non-uniform sources is:
\begin{align}
f_{o1,\infty} &= \frac{S}{\nu_c} \left( 1 + 3 \varepsilon + \left(\frac{\varepsilon}{\alpha}\right)^2 \kappa \frac{2}{3+\left(\frac{\varepsilon}{\alpha}\right)^2}  \right),\label{eq: five equation analytic solution OAP localized source, distributed sink trapped region}\\[5pt]
f_{o2,\infty} &= \frac{S}{\nu_c} \left( 1 + 3 \varepsilon + \left(\frac{\varepsilon}{\alpha}\right) \kappa \frac{3 - \left(\frac{\varepsilon}{\alpha}\right)}{3 + \left(\frac{\varepsilon}{\alpha}\right)^2} \right),\\[5pt]
f_{o3,\infty} &= \frac{S}{\nu_c} \left( 1 + 3 \varepsilon - \left(\frac{\varepsilon}{\alpha}\right) \kappa \frac{3 + \left(\frac{\varepsilon}{\alpha}\right)}{3 + \left(\frac{\varepsilon}{\alpha}\right)^2} \right) ,\\[5pt]
f_{t1,\infty}, f_{t2,\infty} &= \frac{3S}{\nu_c} \varepsilon, \label{eq: five equation analytic solution OAP localized source, distributed sink expander and loss}
\end{align}
where $\varepsilon=\nu_c/\nu_\parallel$.
(Since there are no evolution equations for $f_{ti}$ in an OAP, we use the equations of the FDP.)
In the $\kappa=0$ uniform source limit, these results reduce to equations \eqref{eq: analytic steady state solution ft distributed source, distributed sink} and \eqref{eq: analytic steady state solution expander loss distributed source, distributed sink}.
A localized source creates a gap (relative differences between the steady states of the various $f_{oi}$) of order $\kappa \epsilon/\alpha$ for  $\epsilon \ll \alpha \le 1 $. 
This shift in the values of $f_{oi}$ is physical in the exact FDP ($\alpha = 1$) but is tiny, being comparable to the standard ${\cal O}(\epsilon)$ next-order corrections to lowest-order orbit averaging.
However, this gap is artificially enhanced by a factor of $1/\alpha$ in the OAP.\footnote{The small value of $\alpha$ in the OAP means that the orbit frequency has been slowed down to $\alpha \nu_\parallel$, so the system isn't averaging over the localized source as effectively as in the FDP.}
The dominant relative error ($E_{\rm rel}$) in the equilibrium solution in the OAP in the limit $\varepsilon \ll \alpha \leq 1$ is
\begin{align}
    E_{\rm rel} = \kappa \frac{\epsilon}{\alpha}
    \label{eq:relerr-kappa}
\end{align}
This gives large errors of order unity for the OAP if $\alpha \sim \epsilon$ is used as in the previous section, where we set $\alpha = \nu_c / (\sqrt{3} \nu_\parallel) = \varepsilon / \sqrt{3}$.

\begin{figure}[t]
    \centering
    \includegraphics[width=\linewidth]{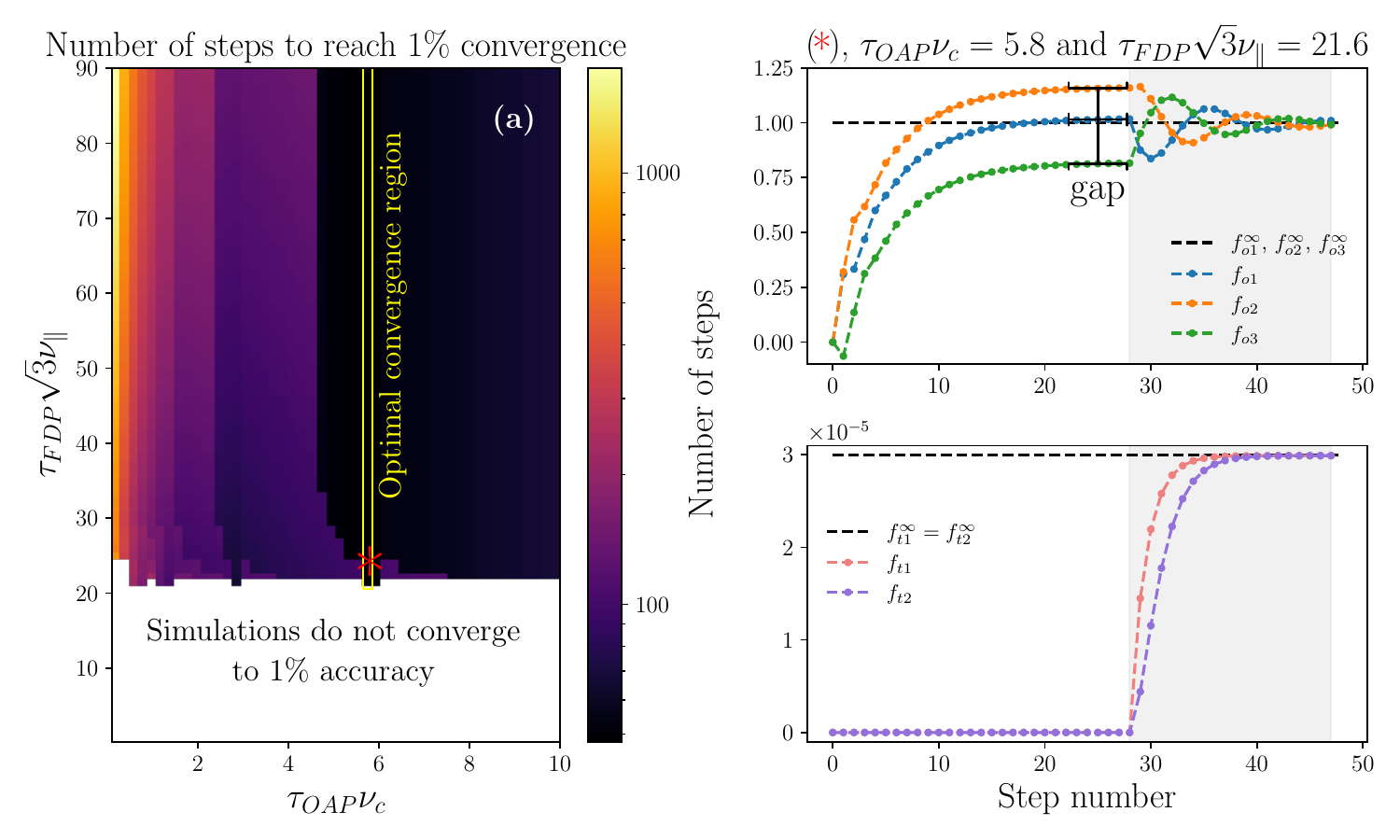}
    \caption{ (a) The number of steps required to reach 1\% convergence in all five variables ($f_{o1}$, $f_{o2}$, $f_{o3}$, $f_{t1}$, $f_{t2}$). In these simulations, $\alpha = 10 \nu_c / (\sqrt{3} \nu_\parallel)$, $\nu_c = 1.0$ (distributed coupling), $S = 1$, $\kappa = 1$ (localized source), $\nu_\parallel = 10^5$, $\nu_{\rm LP} = \sqrt{3} \nu_\parallel$, and simulations end after $T = 100 / \nu_c$ if they haven't converged. A yellow box marks the region that requires the minimum number of time steps (48). Orbiting (b) and transiting (c) solutions for the red asterisk case in (a) are shown on the right. Grey regions show the FDP, while the OAP has a white background. The black dashed lines show the analytical solution for the FDP.}
    \label{fig: number steps to convergence without orbit averaging}
\end{figure}

One approach to reduce this error to a tolerable level is to increase the value of $\alpha$.  Since the speedup scales as $1/\alpha$ and the error scales as $\kappa \epsilon / \alpha$, there is an asymptotic regime with $\epsilon \ll \alpha \ll 1$ where the error is small and the speedup is large.  
For the case with $\epsilon = 10^{-5}$, we could increase $\alpha$ by a factor of 30 to $\alpha = 30 \varepsilon / \sqrt{3}$ to reduce the error to about $6\%$  while still achieving a factor of 1000 speedup.
Also, the degree of non-uniformity of a source in particular real systems might correspond to $\kappa < 1$, further reducing the error.
figure \ref{fig: number steps to convergence without orbit averaging}(b-c) shows a case with $\alpha = 10 \varepsilon/\sqrt{3}$, verifying that the relative errors in the OAP are indeed about $\sqrt{3}/10$, as given by equation \eqref{eq:relerr-kappa}.

The enhancement of the gap (caused by the source localization) in the OAP relative to the FDP means that  oscillations in the solution will be initialized when the simulation switches from an OAP to an FDP. Because the damping rate $\nu_c$ is so small compared to the oscillation frequency $\nu_\parallel$, it can take an extremely long time in the FDP for these oscillations to die away, thus slowing down convergence.  
In some systems, there might be additional damping mechanisms in the FDP, like Landau damping or numerical damping from upwinding, that help these oscillations decay more quickly.

Two strategies involving orbit-averaging are proposed to ameliorate the issues posed by this different gap in the OAP than the FDP. First, a low-pass filter is proposed to be added during the FDP to damp the oscillations induced from the gap, representing an orbit-average in time. 
Second, an explicit orbit average in space may be performed during the OAP. 

\subsubsection{Low-Pass Filter}
\label{sec: low-pass filter}
The gap created during the OAP leads to an oscillatory pulse during the FDP that circulates through the orbit contour. Without intervention, this pulse would damp at the slow collision rate $\nu_c$, and the POA scheme would perform no better than standard integration. To improve the methodology, one can include a low-pass filter during the FDP to damp the advective scale oscillatory modes. During the FDP, we add a new term to the orbiting equations:
\begin{equation}
    \frac{d f_{oi}}{d t} = \ldots - \nu_{\rm LP} \left( f_{oi} - \bar f_{oi} \right)  \label{eq: five equation FDP low-pass filter},
\end{equation}
and we evolve an additional equation
\begin{equation}
    \frac{d \bar f_{oi}}{d t} = \nu_{\rm LP} \left( f_{oi} - \bar f_{oi} \right)  \label{eq: five equation FDP low-pass filter f bar} .
\end{equation}
where $\bar f_{oi}$ is the low-pass filtered, time-averaged $f_{oi}$ and $\nu_{\rm LP}$ is the frequency at which dynamics $\omega > \nu_{\rm LP}$ are damped out. We suggest setting $\nu_{\rm LP} = \omega_\parallel$ to damp oscillations that occur at frequencies higher than the orbit frequency. 
In the results shown here, $\nu_{\rm LP}$ terms are ignored during the OAP, and $\bar{f}_{oi}$ is set to $f_{oi}$ at the beginning of the next FDP.
Strictly speaking, one needs to add $2 \nu_{\rm LP}$ to the denominator of equation \eqref{eq:RK4-dt}, but we remain within the RK4 stability region without it for the choice of $\nu_{\rm LP}=\omega_\parallel$.

To demonstrate the effect of the low-pass filter, a convergence study is performed to show that high rates of convergence can be reached, even when a large gap is introduced during the OAP. Since the gap $\propto \varepsilon \kappa / \alpha$, a larger $\alpha$ is used of $\alpha = 10 \nu_c / (\sqrt{3} \nu_\parallel)$ so that the gap is not order unity. The five-equation model is defined as converged when all five variables ($f_{o1}$, $f_{o2}$, $f_{o3}$, $f_{t1}$, $f_{t2}$) are within $1\%$ of their analytic value. Simulations that do not converge are presented as white cells. Figure \ref{fig: number steps to convergence without orbit averaging}a shows the number of steps the scheme must take to reach convergence, scanning the amount of time spent in the OAP and the FDP. The yellow box on the left plot bounds the region with the minimum number of steps to reach convergence, as 48 RK4 steps. The simulation with the red asterisk is shown on the right to visualize the evolution of $f$. The optimal simulation runs by performing a single OAP that runs for many collision times until it has reached a steady state. In this simulation, it occurs after 28 time steps. The 29th and later steps are determined by solving the FDP equations, including the low-pass filter. 

Although the gap between $f_{o1}$ and $f_{o3}$ is $0.4$ during the OAP, the low-pass filter damps the oscillations, and the solution converges to the orbit-average of $f_o$.
The yellow box continues vertically on the left plot of figure \ref{fig: number steps to convergence without orbit averaging} because even if the FDP is configured to run for longer, the simulation will stop prematurely when $f$ has converged.
Interestingly, the low-pass filter can be used for all $f$, not just the orbiting particles, demonstrating the robustness of the low-pass filter.
A beneficial side-effect of the low-pass filter is that it will damp any spurious waves in the system that occur at this scale. The POA algorithm is intended for simulations targeting a steady state without wave propagation, so the low-pass filter helps condition the problem to this regime by removing all waves faster than the transit time. 
The time spent in the FDP increased compared to figure \ref{fig: number of steps until fiveq converged} from $13.5 /  \sqrt{3} \nu_\parallel$ to $21.6 / \sqrt{3} \nu_\parallel$ and more steps were taken in the OAP because of the larger value of $ \alpha$, with the net effect that the POA scheme achieves a speedup of $7,225\times$ compared to direct RK4.

\subsubsection{Numerically Orbit-Averaging the Distribution Function}
\label{sec: orbit averaging the distribution function}
From this examination, it is clear that the issues with a localized source arise due to the gap in the steady state of the OAP. Another strategy to narrow the gap is to force $f$ toward its orbit-average during the orbit-average phase. 
To motivate such forcing, modify equation \eqref{eq: dfdt reduced dynamics frozen fast phase} to
\begin{align}
\frac{\partial f}{\partial t} & = H_{\text{orbit}}\left[\alpha\{\mathcal{H},f\} + (1-\alpha) \{\mathcal{H},f\} + C(f) + S\right] \nonumber \\
 & \approx H_{\text{orbit}}\left[\alpha\{\mathcal{H},f\} - \nu_{\rm avg} (f - \mathring f) + C(f) + S\right].
 \label{eq: five equation orbit average of f forcing term BGK}
\end{align}
where $\mathring f$ is the orbit average of $f$.
Rather than completely neglect the $(1-\alpha)\{\mathcal{H},f\}$ term in the OAP as we originally did, we instead model it as a BGK-type smoothing operator that relaxes $f$ towards its orbit-average at the rate $\nu_{\rm avg}$, where one might choose $\nu_{\rm avg} = (1-\alpha) \Omega_{\rm max}$, where $\Omega_{\rm max}$ is an estimate of the maximum orbit frequency of the system described by $\cal{H}$.

Calculating the orbit average is easy for the 5-equation model, $\mathring f_o = ( \sum_{j=1}^3 f_{oj} )/3$.
In more general systems, implementing the orbit average $\mathring f = \left[\int ds  f(z(s)) / |v_\parallel(s)| \right] / \int ds / |v_\parallel(s)|$ by numerically integrating along trajectories $z(s)$ in phase space might be somewhat complicated, but might be worth the effort in some cases.
\footnote{For some problems, this integral doesn't need to be done to high accuracy, since it will be followed by an FDP that may correct errors in the OAP from direct orbit-averaging.}
Implementing direct orbit-averaging of $f$ might be easier than implementing the analytical orbit-averaged equations for some systems.
Also, the usual lowest order analytical orbit averaging sets $f=0$ in the transit region as $\mathcal{O}(\epsilon)$ smaller than $f$ in the orbit region, while the way we do things here will calculate $f$ in the transit region as well.

Because $\nu_{\rm avg}$ in equation \eqref{eq: five equation orbit average of f forcing term BGK} is very large, it would impose a tiny time step on our explicit method, thus defeating the attempt to use a small $\alpha$ to increase the time step. To get around this, we use an implicit method to advance this term with a simple operator splitting approach.  
If we were doing a first-order explicit method for all of the other terms, then after finding $f^{n+1}_{\rm exp}$ at the future time from all of the explicit terms, we would then do a backward Euler implicit step for the BGK smoothing operator to find the final solution at the future time,
$f^{n+1} = \lambda f^{n+1}_{\rm exp} + (1-\lambda) {\mathring f}^{n+1}_{\rm exp}$, where $\lambda = 1/(1+\nu_{\rm avg} \Delta t)$.  I.e., this is a weighted average of the explicit solution for $f^{n+1}$ and its orbit-average.  
For typical cases, $\lambda \ll 1$ and one gets almost the same results as the asymptotic orbit-averaging limit $\lambda=0$, i.e., just replacing the explicit solution with its orbit average after every explicit step.
We use an RK4 algorithm for the explicit terms, and it is important to apply this implicit BGK smoothing operator after each of the internal 4 steps of RK4 where intermediate values of $f$ are calculated.
If the orbit smoothing was applied only after a full RK4 step, then asymmetric gap errors (as discussed in section \ref{sec: asymmetric source-damping})
can build up in a single RK4 step, so that the mean value after orbit averaging still has errors.

To examine the impact of the BGK averaging term added to force $f_o$ toward its orbit-average, the RMS error is computed between the numerical and analytic solutions
as
$\sqrt{\sum_{i,t}(f_{oi,t} - f_{oi, \infty})^2/3N_t}$, averaging over $N_t$ points in time $t$ in the last $1/3$ of the simulation.

\begin{figure}[t]
    \centering
    \includegraphics[width=0.7\linewidth]{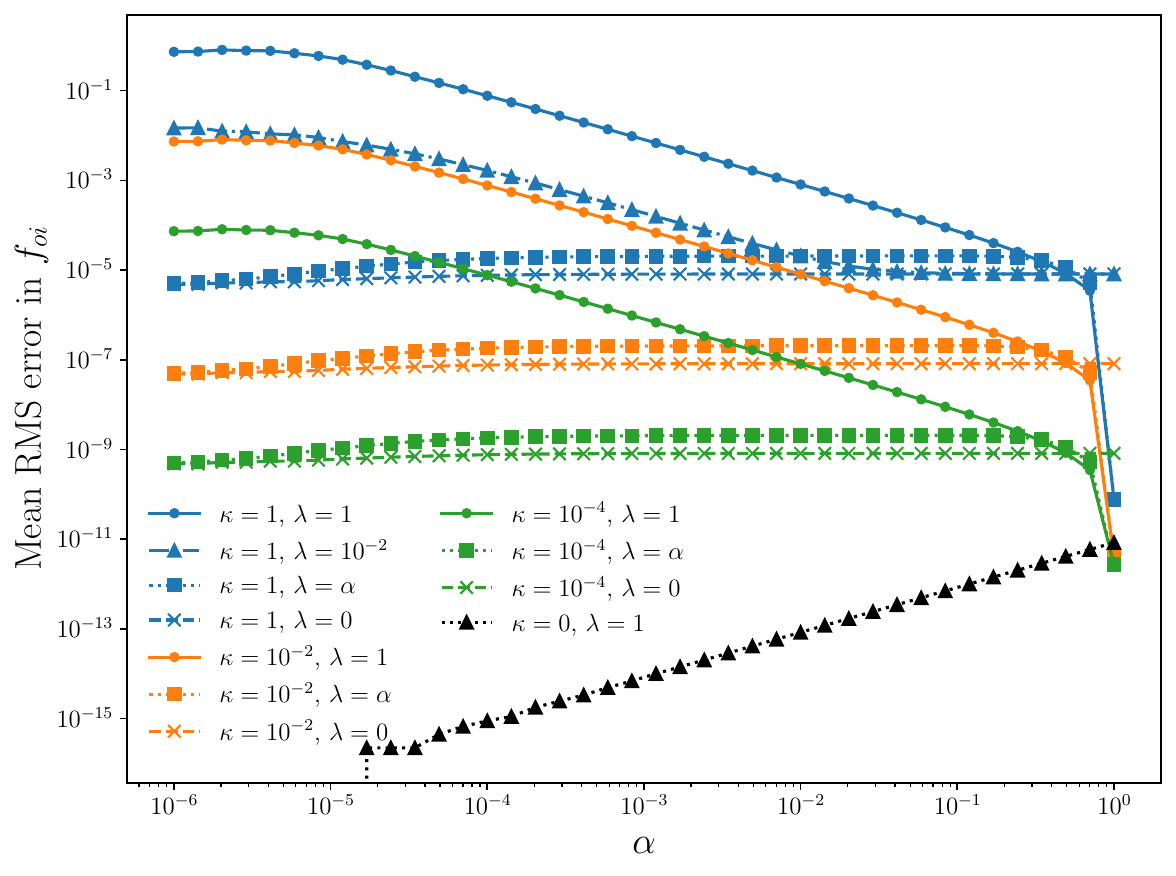}
    \caption{
    RMS error of $f_{o}$ vs. $\alpha$ for different values of parameters ($\kappa=1$ is strongly localized source, $\kappa=0$ is uniform source, $\lambda=1$ is no orbit averaging, $\lambda=0$ is complete orbit averaging after each explicit substep).
    Without orbit-averaging, errors increase as $\alpha$ is lowered, but the orbit-averaging of $f$ mitigates these errors. 
    Other parameters used are $\nu_c = 1.0$, $S = 1$, $\nu_\parallel = 10^5$, $\tau_{\rm OAP} \nu_c = 5.5$, $\tau_{\rm FDP} \nu_\parallel = 14/\sqrt{3}$ and simulations end after $100 \nu_c^{-1}$. $\alpha$ is scanned from $1$ to $10^{-6}$ with a resolution of 100 points. Statistics are recorded in the last $33\nu_c^{-1}$ of the simulation.}
    \label{fig: pareto front error plot}
\end{figure}

Figure \ref{fig: pareto front error plot} shows RMS error for simulations using a parameter set of $S = 1$, $\nu_c = 1$, and $\nu_\parallel = 10^5$. This figure quantifies both the impact of $\kappa$ (controlling source locality), as well as $\lambda$ (strength of orbit averaging), on the error in the dynamics. 
It's clear that each factor of $10$ reduction in $\kappa$ lowers the error by a factor of $10$, indicating a linear relationship. 
The observation that the RMS error is linear on a log-log scale indicates a power law relationship with $\alpha^{-1}$. 
Furthermore, the error appears to be linear with $\lambda$ as well, as shown by the line with $\kappa=1,\,\lambda = 0.01$. 
Most of the results in this figure (except for the $\kappa=0$ case discussed below) can be summarized approximately as $E_{\rm rel} = \lambda \kappa \epsilon / \alpha$ (consistent with equation \eqref{eq:relerr-kappa} modified to include error reduction by the orbit smoothing parameter $\lambda$) and then limited to lie in the range $\kappa \epsilon < E_{\rm rel} < \kappa$.  
The lower limit $\kappa \epsilon$ is set by the minimum error that can be expected from a standard orbit-averaged theory (due to the finite variation of the exact solution from its bounce average).
The upper limit $\kappa$ is for an order unity error relative to the degree of non-uniformity $\kappa$ in the source.

The case in black where $\kappa = 0, \lambda=1$ indicates the case discussed in section \ref{sec: subsection distributed source and sink} with a distributed source and no orbit-averaging of the distribution function. This case produces an RMS error that increases with $\alpha$ due to floating-point precision arithmetic errors. When $\alpha=1$, the system needs many small time steps to reach a time of $100\nu_c$. The small time steps mean that the $f_o \approx \mathcal{O}(1)$ is being added to a small number to adjust. Thus, the small changes are the round-off error, and the solution can only get within a margin of the steady-state. $\alpha < 1$ means that larger time steps are being taken during the OAP, so the floating-point arithmetic errors become narrower. 
For $\alpha < \varepsilon = \nu_c/\nu_\parallel= 10^{-5}$, the RMS error for the $\kappa = 0, \lambda = 1$ case is identically zero, so the dots do not appear on a log-log scale.

Any $\kappa > 0,\, \alpha < 1$ demonstrates some RMS error at least of order $\kappa \epsilon$.  The POA method is not intended to accurately calculate the physical gap, which is of order $\kappa \epsilon$, which is set to 0 in standard analytical orbit averaging at lowest order.
I.e., the POA method can't do better than relative errors of order $\kappa \epsilon$ at a single point.
But the POA method can calculate $f_{ti}$ in the transit regions fairly well, even though $f$ in the transit region is of order $\epsilon$ smaller than $f$ in the orbit region (unlike lowest order analytical orbit averaging which usually sets $f=0$ in the orbit region).

\subsubsection{Discussion}
The introduction of a localized source significantly modifies the steady-state behavior of the five-equation model compared to the distributed source case. Most notably, localization creates a gap between the distribution function at various points along the orbit contour. This gap scales as $\kappa \varepsilon$ during the FDP, but as $\kappa \varepsilon / \alpha$ during the OAP, so the speedup factor $\sim 1/\alpha$ comes at the expense of increased error. 
Consequently, the discrepancy between the two phases grows as $\alpha$ decreases, undermining one of the principal goals of the POA scheme: ensuring that OAPs track the long-time dynamics of the full system.

To draw a connection to physics, realistic mirror fueling schemes—such as neutral beam injection—are highly localized, so a distributed source is not representative of experimental conditions. The mismatch between FDP and OAP steady states reflects the advective slowdown during the OAP, leading to a buildup of information without further modification. Left untreated, the induced oscillatory pulse circulates the orbit contour and damps only at the collision rate, eliminating the speedup benefits of the POA scheme.

Two remedies were therefore proposed. First, a low-pass filter damps advective-scale oscillations during the FDP, effectively orbit-averaging in time. This modification accelerates the damping rate of the induced oscillations from the localized source that otherwise stalls convergence. Numerical tests confirm that the five-equation model with a localized source can converge to the correct steady-state solution with minimal additional computational overhead beyond evolving an auxiliary averaged variable. Second, a BGK-type operator was introduced during the OAP to draw the distribution function toward its orbit-average, orbit-averaging in phase-space. This approach directly counteracts the gap, narrowing it to a proposed scaling of $\kappa \varepsilon \lambda / \alpha$. 
At first it might seem that this term wouldn't be helpful because it would have a very short CFL constraint on the time step, but this is avoided with an implicit time advance for it.
An advantage of using the orbit average of $f$ is that the FDP converges to the analytic answer more quickly than the low-pass filter, which requires the FDP to run roughly 2 times longer. 

Although it was trivial here, numerically computing the orbit-average of a function can be complicated in some systems; thus, it is suggested to start with the simpler versions of POA first without direct orbit averaging (where $\lambda=1$).
One can investigate if a sufficiently small error $\sim \kappa \varepsilon / \alpha$ can be achieved by raising $\alpha$ some, while still getting adequate speedup, which scales as $1/\alpha$.

\subsection{Localized Source, Localized Orbit/Transit Coupling}

More challenging cases are systems where both the sources and the sinks (orbit-transit coupling in our model equations) are localized (and localized in asymmetric ways) instead of being distributed uniformly around an orbit.
An example of this in our model equations could be motivated in figure \ref{fig: schematic of bouncing points} by noting that $f_{o1}$ and $f_{t1}$ are much closer together and so might be more strongly coupled by collisional diffusion than $f_{o2}$ and $f_{o3}$ are coupled to $f_{t1}$.\footnote{There are transit loss orbits symmetrically below the $v=0$ plane that  $f_{o2}$ and $f_{03}$ might be more strongly coupled to by diffusion in some systems, but there are still various ways in which physical systems can have losses that are asymmetrically localized relative to sources, like local neutral gas densities, physical limiters inserted into the edge of a plasmas, etc.}
In terms of our model equations \ref{eq: first statement of five equation model eq 1}-\ref{eq: first statement of five equation model eq 5}, this type of system with asymmetrically localized sources and sinks could be represented by losses localized to $f_{o1}$ (with $\nu_{o2} = \nu_{o3} = 0$, and the sources localized to $f_{o3}$, with $S_1 = S_2 = 0$.  We will discuss ways to handle the asymmetric localized case below, but we start with the easier symmetric localized case.

\subsubsection{Orbit/Transit Coupling and Source with Symmetric Localization}
To consider a case where the source and sink are localized at the same location (and so are symmetric with respect to each other),
in equations \eqref{eq: first statement of five equation model eq 1 orbit average phase} - \eqref{eq: first statement of five equation model eq 5 orbit average phase} set $S_1 = 3S$, $\nu_{c1}=3\nu_c$, and $S_2=S_3=\nu_{c2}=\nu_{c3}=0$.
\footnote{One could get similar results to those in this section if the source and sink terms were localized at exactly opposite sides of an orbit, but because we have an odd number of points per orbit, we can't reproduce that case in these model equations.}
During the FDP phase, the coefficients $\alpha = H_{\rm orbit} = 1$, while during the ODP phase, $H_{\rm orbit} = 0$ and $\alpha$ is specified below.
We chose the localized damping rate to be stronger to give the same oribt-average equilibrium result, so the CFL time step restriction differs from the distributed case and is now
\begin{equation}
\Delta t  = \frac{2.3}{3\nu_c + \alpha\sqrt{3} \nu_\parallel}.
\end{equation}
Choosing $\alpha$ so that the frequency and damping are comparable gives
$\alpha = \sqrt{3}\nu_c / \nu_\parallel$. 
The FDP system of equations yields an identical equilibrium solution to equation \eqref{eq: analytic steady state solution ft distributed source, distributed sink}-\eqref{eq: analytic steady state solution expander loss distributed source, distributed sink}. Both the OAP and the FDP have the same solution, since the factor of $\alpha$ divides out.
\begin{align}
    f_{o1,\infty} = f_{o2,\infty} = f_{o3,\infty} &= \frac{S}{\nu_c}\left(1 + 3 \varepsilon\right), \\
    f_{t1,\infty} = f_{t2,\infty} &= \frac{3S}{\nu_c}\varepsilon . 
\end{align}

\begin{figure}[t]
    \centering
    \includegraphics[width=\linewidth]{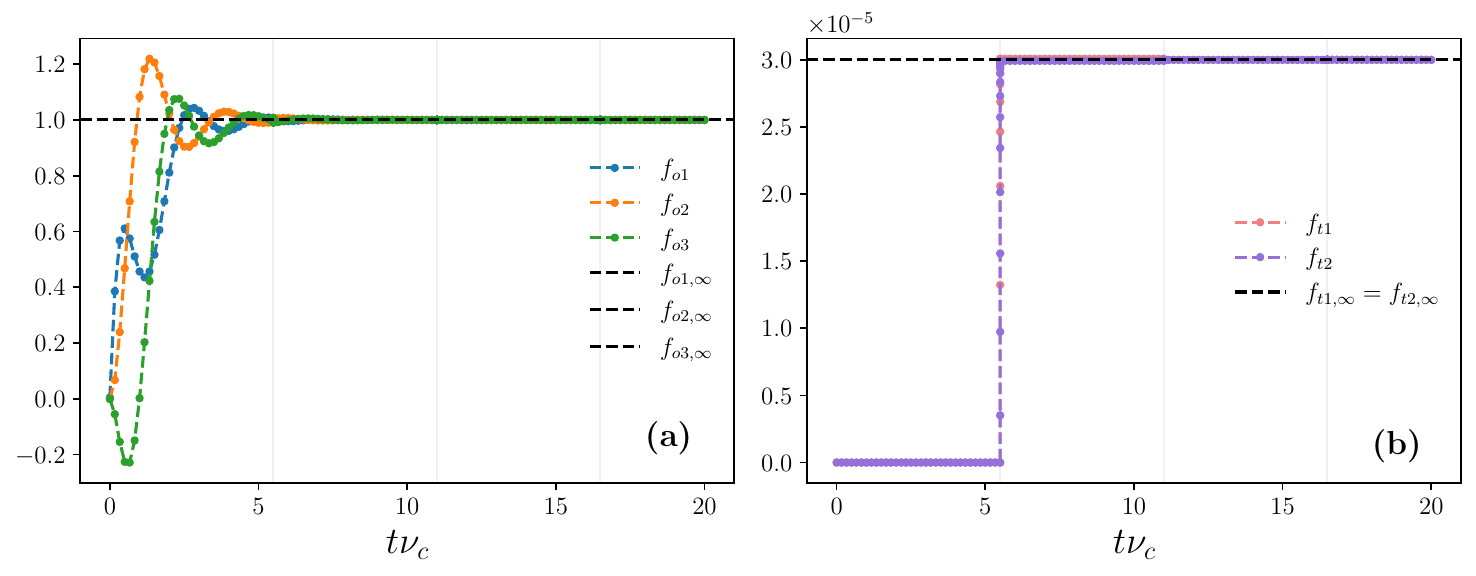}
    \caption{Orbiting (a) and transiting (b) solutions for the case where the orbit/transit coupling and source are co-localized to $f_{o1}$, using parameters $S=1.0$, $\nu_c=1.0$, $\nu_\parallel=10^5$, $\alpha = \sqrt{3}\nu_c/\nu_\parallel$, $\tau_{\rm OAP} \nu_c = 5.5$, $\tau_{\rm FDP} \nu_\parallel = 4.7 \sqrt{3}$.  Grey regions show the FDP, while the OAP has a white background. The simulation runs for $20\nu_c^{-1}$, then is stopped. After 76 steps, all variables reach a relative error compared to the analytic solution $\approx 3\times 10^{-9}$.}
    \label{fig: colocalized sinks and source demonstration}
\end{figure}

One would expect that the localized source and the localized orbit/transit coupling cancel each other out during the OAP, yielding a steady-state solution. Solving the OAP equations leads to oscillations around the analytical solution, but they damp slowly through collisions and converge to the analytical solution. Furthermore, $f_{t2}$ and $f_{t1}$ converge rapidly. To demonstrate this behavior, figure \ref{fig: colocalized sinks and source demonstration} shows a simulation that has run for a few POA cycles. $f_o$ initially oscillates with the suppressed advective frequency $\alpha\sqrt{3} \nu_\parallel$, which decay rapidly. Although these oscillations are not seen in a direct simulation ($\alpha=1$), they are not detrimental to the POA scheme converging on the correct answer. When the system is run longer, ending at $50\nu_c^{-1}$, the error between the analytic expressions and the POA scheme is machine precision zero.

In this symmetric localized case, the system convergences rapidly and doesn't need a low-pass filter or numerical bounce-averaging.
This is because the orbit/transit coupling immediately compensates for the source since it is within the same equation. There is no timescale described by the problem that slows down the dumping of information from the source directly into the transit region. Since the steady-state equation is identical during the FDP and OAP, this regime is similar to that studied in section \ref{sec: subsection distributed source and sink}. The OAP does not induce a gap in the solution and converges quickly to the correct value. 

\subsubsection{Orbit/Transit Couplings and Sources with Asymmetric Localization}
\label{sec: asymmetric source-damping}
The most challenging case is where the orbit/transit coupling and source are localized at different locations (and in an asymmetric way, not at exactly opposite locations).
This section explores a localized orbit/transit coupling in $f_{o1}$, but the source is placed at $f_{o2}$. Equations \eqref{eq: first statement of five equation model eq 1 orbit average phase} - \eqref{eq: first statement of five equation model eq 5 orbit average phase} are therefore modified such that  $S_2 = 3S$, $\nu_{c1}=3\nu_c$, and $S_1=S_3=\nu_{c2}=\nu_{c3}=0$.
The analytic equilibrium is
\begin{align}
    f_{o1,\infty} = f_{o2,\infty} &= \frac{S}{\nu_c}\left(1 + 3 \varepsilon\right), \\
    f_{o3,\infty} &= \frac{S}{\nu_c}\left(1 + 3 \varepsilon + 3\frac{\varepsilon}{\alpha}\right), \\
    f_{t1,\infty} = f_{t2,\infty} &= \frac{3S}{\nu_c}\varepsilon . 
\end{align}
The separation between the localized source and orbit/transit coupling exposes a gap. During the FDP, the gap has a width $3S\varepsilon/\nu_c$. The gap is affected by the OAP and widened to be $3S\varepsilon/(\alpha\nu_c)$, putting an $\alpha$ in the denominator. In these equations, it takes time for $f$ to advect from $f_{o2}$ all the way to $f_{o1}$, which is even longer during the OAP by a factor $\alpha$.

\label{sec: counterpositional localized source and sinks}
\begin{figure}[t]
    \centering
    \includegraphics[width=\linewidth]{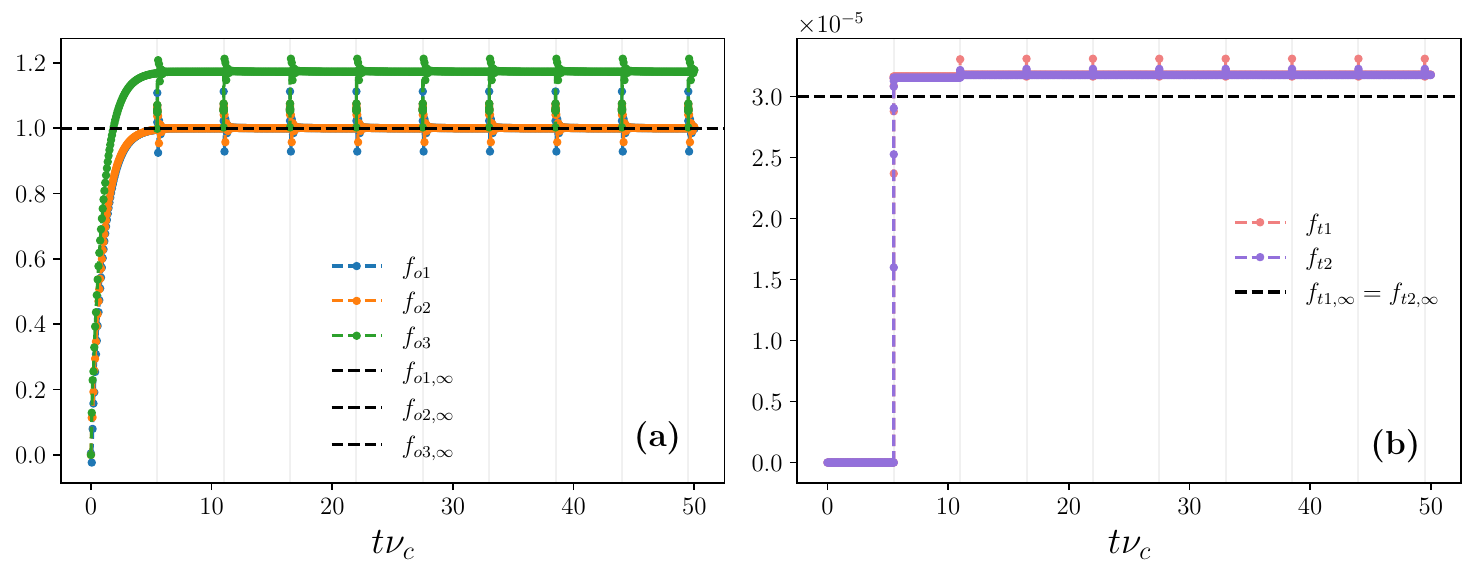}
    \caption{Orbiting (a) and transiting (b) solutions for the case where the orbit/transit coupling is localized to $f_{o1}$ and the source is localized to $f_{o2}$, using parameters $S=1.0$, $\nu_c=1.0$, $\nu_\parallel=10^5$, $\alpha = 10\sqrt{3}\nu_c/\nu_\parallel$, $\tau_{\rm OAP} \nu_c = 5.5$, $\tau_{\rm FDP} \nu_\parallel = 4.7 \sqrt{3}$.  Grey regions show the FDP, while the OAP has a white background. No time-filtering or direct orbit averaging is used. The simulation runs for $50\nu_c^{-1}$, then is stopped. The black dashed lines show the steady-state solution for the FDP.}
    \label{fig: counterlocalized sinks and source demonstration}
\end{figure}

An example POA scheme simulation is shown in Figure \ref{fig: counterlocalized sinks and source demonstration}. 
A value of $\alpha$ that is $10\times$ larger than the value used in the last section of $\sqrt{3}\nu_c/\nu_\parallel$ is used to narrow the gap so that it is not order unity. It is seen that during the OAP, $f_{o3}$ rises above $f_{o1},f_{o2}$ by $20\%$, which induces oscillations during the FDP. Furthermore, the average value of these oscillations is shifted higher than the steady-state solution for the FDP, which leads $f_{t1},f_{t2}$ to take a value higher than the analytic steady state.

\begin{figure}[t]
    \centering
    \includegraphics[width=\linewidth]{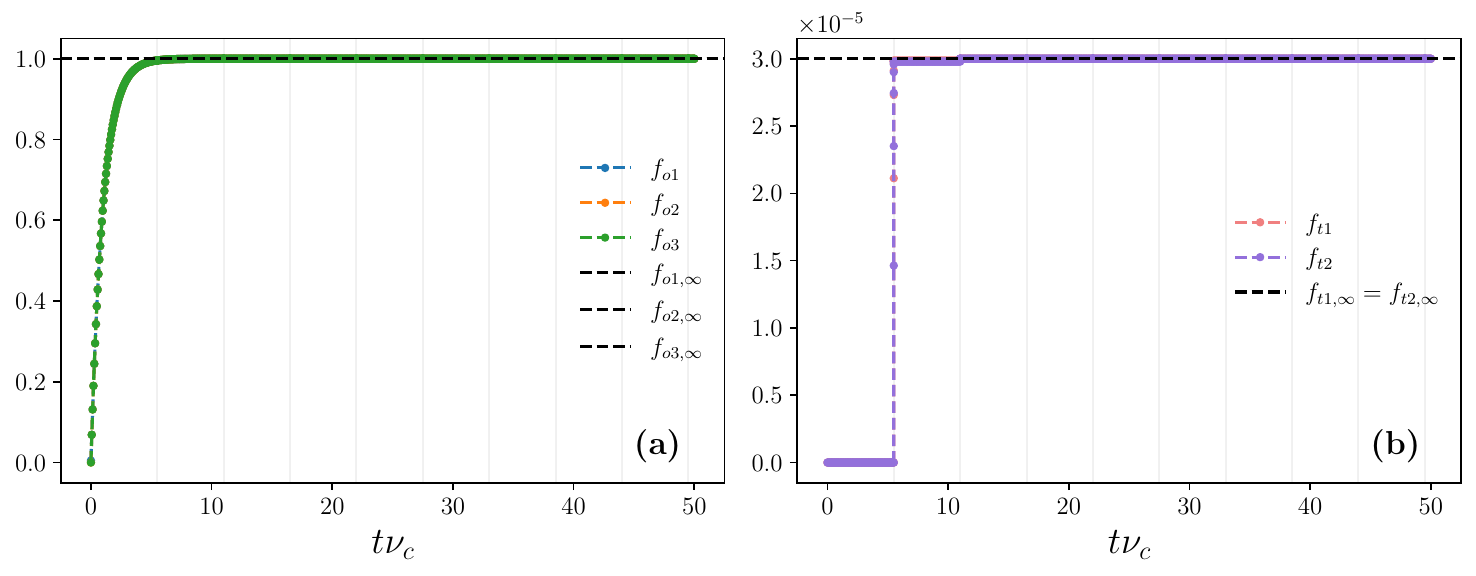}
    \caption{The same case explored in figure \ref{fig: counterlocalized sinks and source demonstration}, but applying explicit orbit averaging ($\lambda=0$) during the OAP. }
    \label{fig: counterlocalized sinks and source demonstration with orbit averaging}
\end{figure}

The one-sided nature of the gap in this problem means that if one were to apply a low-pass filter during the FDP 
the result would not converge to the analytic solution 
(not on a short time scale of the filter, one would have to run for a very long $1/\nu_c$ time in the FDP to reach equilibrium, but then there would be no computational speedup).
The time-filtering technique in section \ref{sec: low-pass filter} was effective because the gap was symmetric, meaning the orbit-transit coupling was centered at the correct value. The gap shifted $f_o$ such that half went higher and half went lower. However, here the gap is one-sided, causing the $f_{t1}$ to shift up. Another way to understand this is that the slowdown of advection creates a pile-up of information in $f_{o2}$, increasing the total amount of information in $f$. 

This problem can be fixed by using the direct orbit averaging during the OAP, as described in section ~\ref{sec: orbit averaging the distribution function}.
As shown in figure \ref{fig: counterlocalized sinks and source demonstration with orbit averaging}, when direct orbit averaging is used ($\lambda=0$), the gap during the OAP vanishes, and the solution converges. The $f_{o3}$ in figure \ref{fig: counterlocalized sinks and source demonstration} has an difference of $0.179$ with respect to the analytic value, however $f_{o3}$ in figure \ref{fig: counterlocalized sinks and source demonstration with orbit averaging} has a difference of only $2.99 \times 10^{-5}$, an $\mathcal O(\varepsilon)$ error (the same as analytical orbit averaging).

In the current 5-equation model, there are only 3 points around an orbit, so one can't investigate the effects of the relative position of the source and sink in more detail.
To enhance the resolution in phase-space angle, consider a continuum PDE variant of the five-equation model. Consider the continuum advection model only in the orbit region with a source and orbit/transit coupling localized at two locations. The advection is at a constant rate $\Omega$ in angular periodic coordinates so that $\theta \in [0,2\pi)$. Collisions are modeled as a simple sink term $-\nu_c f$ and localized at a certain point
\begin{equation}
    \frac{\partial f(\theta,t)}{\partial t} + \alpha \Omega \frac{\partial f}{\partial \theta} =  - \nu_c f \delta(\theta - \theta_{\rm sink}) + S \,\delta(\theta)\label{eq: five-eq continuum model pde localized source and sink}
\end{equation}
The problem may be solved for its steady state solution. Between the delta functions, the right-hand side is zero, so the solution is $f=\rm constant$. The jump conditions show that the jump at zero is $\Delta f(0) = S/\Omega$ and at the sink $\Delta f(\theta_{\rm sink}) = (2 \Omega + \nu_c)/(2 \Omega - \nu_c)$. The upper and lower values of $f$ are the same in both jumps, and they are solved. The solution is
\begin{align}
    f^+_{\rm OAP} = \frac{S}{\nu_c} \left( 1 + \frac{\nu_c}{2 \alpha \Omega} \right), \label{eq: continuum oap gap asymmetric localization top}\\
    f^-_{\rm OAP} = \frac{S}{\nu_c}  \left( 1- \frac{\nu_c}{2 \alpha \Omega} \right) \label{eq: continuum oap gap asymmetric localization bottom}. 
\end{align}
where $f^+$ is the upper solution and $f^-$ is the lower solution. The upper solution has a domain downstream from the source. Each plateau has a width $\theta_{\rm sink}$ and $2 \pi - \theta_{\rm sink}$. From equations \eqref{eq: continuum oap gap asymmetric localization top}-\eqref{eq: continuum oap gap asymmetric localization bottom}, it is clear that the orbit-average converges to $S/\nu_c$ if and only if the localized orbit/transit coupling and source are placed exactly opposite each other, so $\theta_{\rm sink} = \pi$. When the localized orbit/transit coupling and source are placed counter-symmetrically, the OAP and FDP have the same orbit-average, meaning that orbit-average techniques are appropriate to narrow the gap. When the widths of the regions are different, such as in the three-point five-equation model or in physical systems where the sources and sinks are not symmetrically located relative to each other, the orbit-average is not the same in each phase, and time averaging techniques in the FDP won't help accelerate convergence to the proper long-time equilibrium solution. 

\subsubsection{Discussion}
Localizing both the source and the orbit/transit coupling reveals two distinct behaviors that are important for practical POA implementations. When the orbit/transit coupling is co-positional with the source, the local injection and immediate removal essentially cancel out on the orbit contour, so the OAP and FDP share the same equilibrium, and the POA scheme remains accurate and rapidly convergent. In that regime, the OAP does not introduce a gap. Additional orbit averaging strategies are not necessary.

By contrast, systems where the orbit/transit coupling is localized at a different position than a localized source (the counter-positional case) produces a pronounced, one-sided gap (for the general asymmetric case, unless the source and sink are exactly antisymmetric): advection must carry particles from the injection site to the spatially displaced orbit/transit coupling, and this transport is slowed during the OAP by the factor $\alpha$, producing a pileup (in $f_{o2}$). This asymmetry is qualitatively different from the symmetric gaps produced by localized sources with a distributed orbit/transit coupling. The most accurate and direct way to handle this case is to use the direct orbit-averaging approach of section \ref{sec: orbit averaging the distribution function}.
This is numerically similar to analytical orbit-averaging, but goes beyond the usual lowest-order orbit-averaging to also calculate $f$ in the transit-loss regions.

If direct orbit-averaging of $f$ hasn't been implemented, there are two more approaches to consider.
Increasing $\alpha$ above $\epsilon$ can mitigate the effect by narrowing the gap, though it sacrifices some speedup.
The relative error caused by asymmetry in the gap in the OAP scales as $\delta \epsilon / \alpha$, where $\delta$ is a measure of the degree of asymmetry in the sources and sinks ($\delta$ is order unity for strongly asymmetric localized sources and sinks).
If $\epsilon = \nu_c / \Omega = 10^{-5}$ (as in some systems we are interested in), and if 3\% accuracy is sufficient, then one can run with $\alpha = 3 \times 10^{-4}$ and still get a factor of $\sim 3000$ speedup. Another approach is to note that the solution from the POA algorithm has the form $f = f_{\rm exact} ( 1 + C \epsilon / \alpha)$, where $C$ is an order unity coefficient.
(This expression keeps terms of order $\epsilon/\alpha$ but neglects terms of order $\epsilon$).  Then one can calculate the POA approximation for two different values of small $\alpha$ and extrapolate to $\alpha =1$ to get a more accurate solution
(this is somewhat like Richardson extrapolation in convergence studies).
\subsection{Distributed Source, Localized Orbit/Transit Coupling}

Lastly, we can construct a model with a localized source and distributed orbit/transit coupling. We localized the orbit/transit coupling to $f_{o1}$ to mirror figure \ref{fig: schematic of bouncing points}; however, moving the orbit/transit coupling to other equations induces a phase shift in the final answer. Here, we reconfigure equations \eqref{eq: first statement of five equation model eq 1 orbit average phase} - \eqref{eq: first statement of five equation model eq 5 orbit average phase} by setting $S_1=S_2=S_3=S$, $\nu_{c1}=3\nu_c$, and $\nu_{c2}=\nu_{c3}=0$.
The steady state solution for general $\alpha$ is
\begin{align}
    f_{o1,\infty} &= \frac{S}{\nu_c}\left(1 + 3 \varepsilon\right), \\
    f_{o2,\infty} &= \frac{S}{\nu_c}\left(1 + 3 \varepsilon - \frac{\varepsilon}{\alpha}\right), \\
    f_{o3,\infty} &= \frac{S}{\nu_c}\left(1 + 3 \varepsilon + \frac{\varepsilon}{\alpha}\right)\\
    f_{t1,\infty} = f_{t2,\infty} &= \frac{3S}{\nu_c}\varepsilon .
\end{align}
The FDP steady state is given by these equations with $\alpha=1$.
It is clear from these equations that the gap is centered at $(S/\nu_c) \left(1 + 3 \varepsilon\right)$, independent of which phase the system is in. 

\begin{figure}[t]
    \centering
    \includegraphics[width=\linewidth]{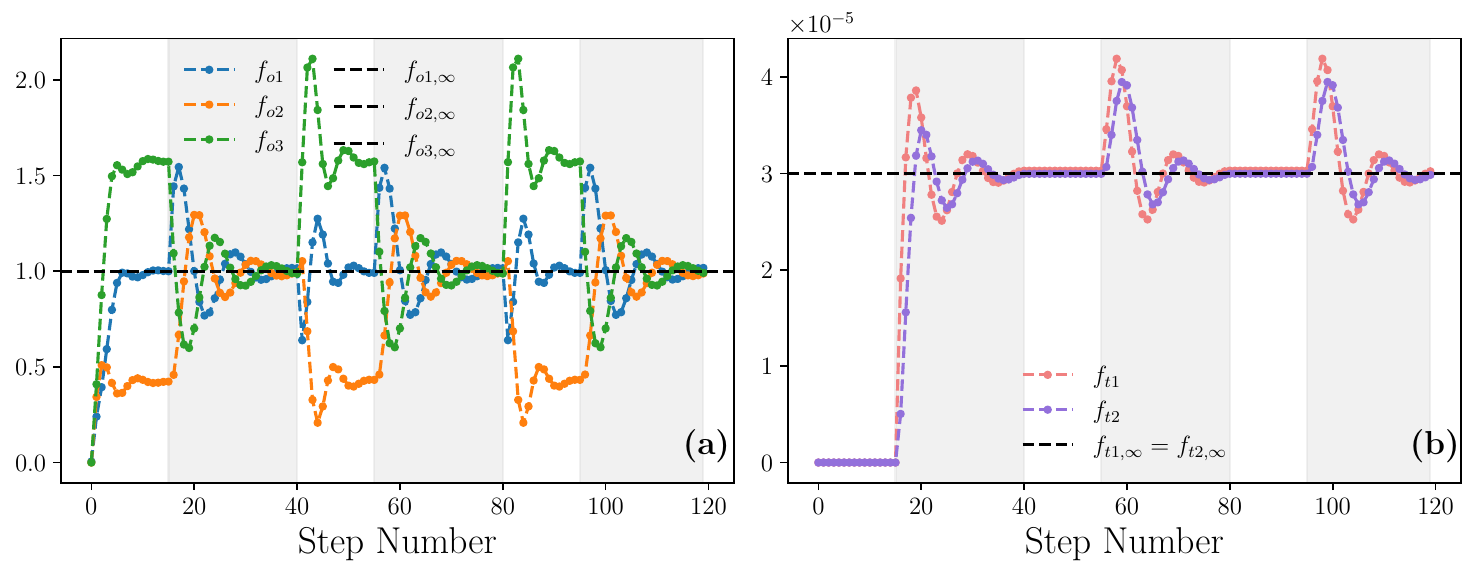}
    \caption{Orbiting (a) and transiting (b) solutions for the case with a distributed source and the orbit/transit coupling localized to $f_{o1}$, using parameters $S=1.0$, $\nu_c=1.0$, $\nu_\parallel=10^5$, $\alpha = \sqrt{3}\nu_c/\nu_\parallel$, $\tau_{\rm OAP} \nu_c = 5.5 $, $\tau_{\rm FDP} \nu_\parallel = 4.7  \sqrt{3}$. Three cycles of the POA scheme are shown. Time filtering is used ($\nu_{LP} = \sqrt{3}\nu_\parallel$), but no orbit average techniques are employed.  Grey regions show the FDP, while the OAP has a white background. The black dashed lines show the steady state of the FDP.}
    \label{fig: demonstration localized sink, distributed source}
\end{figure}

An example evolution of the five-equation model with a distributed source and localized orbit/transit coupling is shown in Figure \ref{fig: demonstration localized sink, distributed source}, with time filtering as described in section \ref{sec: low-pass filter}. 
The OAP yields a gap much larger than the intended solution during the FDP. The low-pass filter is applied over three cycles to demonstrate that the solution decays toward the appropriate value. The third POA phase ends with all variables converging to $1\%$ of the equilibrium of the FDP. When direct orbit averaging is applied, the results are similar to figure \ref{fig: counterlocalized sinks and source demonstration with orbit averaging}, where errors are $\mathcal{O}(\varepsilon)$, as the gap during the OAP is directly opposed. Oscillations of $f_{t1}$ and $f_{t2}$, demonstrating how the localized orbit/transit coupling couples the oscillations in $f_{o1}$ to $f_{t1}$, whereas the distributed orbit/transit coupling averages over the advective oscillations in the orbiting region. 

\subsubsection{Discussion}
A distributed source with localized coupling to the transit region produces a symmetric gap, ameliorating the issues posed by section \ref{sec: counterpositional localized source and sinks}. Analytically, the gap in the steady state remains centered at the same mean $(S/(\nu_c) (1+3\varepsilon))$ during the FDP and the OAP. Still, the OAP shifts orbiting cells symmetrically by $\pm \varepsilon/\alpha$. The localization of the orbit/transit coupling couples the advective oscillations in the orbiting region directly into the transit/exhaust populations. 
These results suggest another strategy to address the issues posed by section \ref{sec: counterpositional localized source and sinks} for systems that have asymmetric localized sources and sinks. 
In some cases it might be easier to orbit-average the source term by itself (such as if it is time-independent or can be expressed in terms of constants of the motion) rather than calculate the orbit average of $f$ after every time step.  
Even though the sink terms are still localized, the gap is now symmetric, so applying low-pass time filtering can damp the oscillations in an FDP to the correct equilibrium value.

\section{Conclusion}
\label{sec: conclusion}

This article presented a new multiscale explicit time integration algorithm to drastically accelerate numerical modeling of systems with identifiable, separated regions of slow orbiting and fast transiting dynamics.
This pseudo orbit-averaging (POA) algorithm is a numerical analog to analytical orbit averaging.  
An example of such a system is the kinetic description of a weakly collisional magnetized plasma, for which orbit-averaged (or bounce-averaged) equations are sometimes used.
The new POA method presented here consists of alternating between a full dynamics phase (FDP), where the original model is solved, and an orbit-averaged phase (OAP), where transiting regions are frozen and fast dynamics within the orbiting region are slowed down.
This is a first-of-its-kind method for computing steady-state equilibria for relevant systems, and its simplicity makes it an attractive alternative (to complex and sometimes expensive implicit methods) that can be trivially implemented in existing codes.

The algorithm's efficacy is demonstrated through two complementary reduced models. Using the one-dimensional source-diffusion-sink PDE model, it is explained how the passing dynamics influence the orbit region. It is shown, in continuum, that the algorithm agrees with the analytic equilibrium for both the orbiting and transiting parts of the problem, introducing no numerical artifacts. A five-equation ODE model is examined, where three equations model orbital advection and two model transiting trajectories with rapid losses. Various configurations of the source and orbit/transit coupling are examined, whether they are localized or distributed. The best performing case had a distributed source and orbit/transit coupling, achieving a 
30,000-fold reduction in computational cost (with realistic parameters for a physical problem of interest) while accurately capturing the system's steady-state behavior. %

If there is a strong localization of the source and orbit/transit coupling, this creates a gap (variations in $f$ around an orbit in the equilibrium solution), which widens with small $\alpha$, the factor by which the fast dynamics are slowed down. Several strategies can be employed to mitigate the inconsistency of the solution during the FDP and OAP.
In some cases, a temporal average can be performed using a low-pass filter, which is easy to implement.
If the sources and sinks are asymmetric relative to each other, then the gap can become asymmetric and time filtering is not sufficient.
One option is simply to run with a larger $\alpha$, which in some cases can reduce the error to acceptable levels while still achieving adequate speedup. 
Another option notes that the solution is asymptotically $f = \bar{f} + b /\alpha$, where $\bar{f}$ is the orbit-averaged solution. Then run the algorithm for 2 different values of $\alpha$ to determine the small value of the gap amplitude $b$, and do a Richardson-like extrapolation to the exact limit of $\alpha=1$.
If it is feasible to numerically calculate the orbit average of $f$, then the best approach is to use a BGK-type operator that relaxes the solution to its orbit average during the OAP.

For many problems, the main POA algorithm as described in section \ref{sec:POA-algorithm} can work well without additional time smoothing or numerical averaging. 
It worked well in our tests of it in our full discontinuous Galerkin gyrokinetic code Gkeyll\citep{gkeyllGkeyllDocumentation,Francisquez_Cagas_Shukla_Juno_Hammett_2025} when applying it to calculate 3D collisional plasma equilibrium in a mirror (which we will report on separately). 
In part, this is due to additional damping mechanisms present in the full kinetic code, such as Landau damping and hyperdiffusion from upwind fluxes, that aren't present in the reduced models considered here. It is also because $\alpha$ for that problem is somewhat larger than the $\sim \nu_c/\Omega$ used in test problems here, because of other constraints on the time step.

Future publications will demonstrate the use of the POA algorithm in advanced kinetic plasma simulation frameworks to model experimentally relevant mirror-trapped plasmas. We have already implemented the POA integrator in the gyrokinetic solver of the Gkeyll code, and have computed some gyrokinetic equilibria with a transformative computational speed-up \citep{gkeyllGkeyllDocumentation,Francisquez_Cagas_Shukla_Juno_Hammett_2025}. Longer-term prospects, now enabled by this new capability, include the calculation of mirror equilibria with kinetic electrons, higher-fidelity collision operators, multi-well magnetic geometry, or even neutral interactions.

\section*{Acknowledgments}
This work was supported by the U.S. Department of Energy under contract number DE-AC02-09CH11466 at the Princeton Plasma Physics Laboratory, through a DOE Distinguished Scientist award, the CEDA SciDAC project, and other projects. 
The United States Government retains a non-exclusive, paid-up, irrevocable, worldwide license to publish or reproduce the published form of this manuscript, or allow others to do so, for United States Government purposes. Part of this work was funded by the Department of Energy, Office of Fusion Energy Science FIRE Collaborative: Fusion Neutrons for Integrated Blanket Technology Development Through Advanced Testing and Design, as part of the work to understand HTS mirrors as a neutron source (Award DE-SC0026040).

 During the preparation of this work, the authors used GitHub Copilot, ChatGPT, and Google Gemini in order to generate drafts of research notes, evaluate ideas, and design software. After using this tool/service, the authors reviewed and edited the content as needed and take full responsibility for the content of the published article.

\section*{Data Availability Statement}
The code used to generate the figures in this paper is stored in a public GitHub repository and can be found at \url{https://github.com/Maxwell-Rosen/POA-model-paper-scripts}.

\bibliographystyle{elsarticle-num-names.bst}
\bibliography{bibliography}

\end{document}